\documentclass[onecollarge]{svjour2}       
\smartqed  
\usepackage{graphicx}
\usepackage{xcolor}
\begin{document}

\title{A KPZ Cocktail- Shaken, not stirred...
}
\subtitle{Toasting 30 years of kinetically roughened surfaces}

\author{Timothy Halpin-Healy         \and Kazumasa A. Takeuchi 
}


\institute{T. Halpin-Healy \at
              Physics Department, Barnard College, Columbia University, 3009 Broadway, NY NY 10027 USA \\
              \email{healy@phys.columbia.edu}           
           \and
           K. A. Takeuchi \at
           Department of Physics, University of Tokyo, 7-3-1 Hongo, Bunkyo-ku, Tokyo 113-0033, Japan\\
           \emph{Present address: Department of Physics, Tokyo Institute of Technology, 2-12-1 Ookayama, Meguro-ku, Tokyo 152-8551, Japan.}  \\
}

\date{Received: 18 May 1986 / Accepted: 3 June 2015}

\maketitle

\begin{abstract}
The stochastic partial differential equation proposed nearly three decades ago by Kardar, Parisi and Zhang (KPZ) continues to inspire, intrigue and confound its many admirers. Here, we i) pay debts to heroic predecessors,  ii) highlight additional, experimentally relevant aspects of the recently solved 1+1 KPZ problem, iii) use an expanding substrates formalism to gain access to the 3d radial KPZ equation and, lastly, iv)  examining extremal paths on disordered hierarchical lattices, set our gaze upon the fate of $d$=$\infty$ KPZ.  Clearly, there remains ample unexplored territory within the realm of KPZ and, for the hearty, much work to be done, especially in higher dimensions, where numerical and renormalization group methods are providing a deeper understanding of this iconic equation.
\end{abstract}

\section{In a Nutshell}
\label{intro}

The history of physics has been punctuated at seminal moments by the appearance of certain fundamental equations (and associated models…)  which have vigorously propelled the enterprise forward, serving as an explosively rich departure point, generating a myriad of alternative perspectives, creative insights, surprising connections and, given its sustained impregnability, often remain for many years a sacred object of fascination and obsession to its dedicated disciples.  Recent, obvious suspects in this regard include the quantum mechanical Schr\"odinger equation (equally well, its flipside- Feynman's path integral formulation), or the wonderfully elusive Navier-Stokes equation governing fluid mechanics, by which  may be gleaned the scaling secrets of turbulent flow, dynamically encoded in the whirling eddies drawn by da Vinci centuries ago.  Within the domain of equilibrium statistical physics, the 2d Ising Model, with its tour-de-force algebraic solution by Onsager, followed by combinatoric, graphical, Grassmannian, Monte Carlo, as well as full-blown field-theoretic, scaling, and renormalization group treatments, represents an extraordinary legacy that continues unabated to this very day.  Arguably, a non-equilibrium statistical mechanical analogue to Ising/Onsager is the iconic equation \cite{KPZ} proposed a generation ago by Kardar, Parisi, and Zhang (KPZ); it captures the statistical fluctuations of a {\it kinetically-roughened} scalar height field $h({\bf x},t)$:
$$\partial_th=\nu\nabla^2 h +{1\over 2}\lambda(\nabla h)^2 + \surd D \eta,$$
subject to a i) simple laplacian smoothening mechanism, ii) tilt-dependent local growth velocity,
and, finally, iii) spatiotemporally uncorrelated noise $\eta({\bf x},t)$, which works in competition with the preceding deterministic pieces of this
highly generic, but stubbornly resistant stochastic PDE.  Here, $\nu,\hspace{0.5mm}\lambda\hspace{0.5mm}$ and $D$ are coarse-grained, macroscopic phenomenological parameters, calculable for any given microscopic model via the
technical machinery  of KPZ scaling theory, elucidated by Krug, Meakin and Halpin-Healy (KMHH) \cite{KMHH,S12}, built upon earlier
finite-size scaling ideas \cite{KM}, but predicated ultimately on the central role of the key scaling parameter $\theta= A^{1/\alpha}\lambda$, where $A=D/2\nu,$ and a divergent, parallel correlation length $\xi_\parallel=(\sqrt{A}\lambda t)^{1/z}.$  The 1$^{st}$ KPZ Revolution, 1985-1999, was characterized by exact \cite{HHF,Dhar,Gwa1,Gwa2,KB,DNR,IS88,Dot}, renormalization group \cite{Medina,2loop}, mode-coupling \cite{BKS,MCRG}, and numerical investigations \cite{HH85,MK85,KK89,FT,AF90,FT2,MW94},  whose primary focus was pinning down the dimension-dependent {\it saturation-width} and {\it transient-regime} scaling indices: $\alpha$ and $\beta$=$\frac{\alpha}{z}.$ Galilean invariance, a key underlying symmetry of the KPZ equation was guarantor of the sacred, dimension-{\it independent}
index identity $\alpha + z=2$.   There was, even at this early time, a genuine appreciation of amplitude ratios \cite{HF,LeiHan}, as well as indications  of underlying,  universal KPZ PDFs \cite{KBM,HH91}. Complementary experimental work was scant, but provocative- see early, comprehensive reviews \cite{PM93,HHZ,K97}. 
This epoch closed with the beginnings of Finnish investigations \cite{FFF97} of kinetically-roughened KPZ firelines, key mathematical papers \cite{BG,Gunter}, as well as refreshing nonperturbative \cite{LK} and conformally invariant \cite{Lassig} perspectives, the former inspiring numerical rebuttals \cite{TAP98,JMK98,Roma,Parisi} of a stubborn, battered suggestion \cite{HH90}, revived nevertheless shortly thereafter \cite{FTJ}, that 4+1 might be the upper critical dimension (UCD) of the KPZ problem. Recent numerics \cite{PS12,Parisi13,JMK13,Alves14} appear to have buried this idea, but from the analytical side \cite{glassy,CCDW}, there's lingering suspicion that something nontrivial is afoot near this particular dimension. 

The 2$^{nd}$ KPZ Revolution, commenced in 2000 with the spectacular physical insights of  Pr\"ahofer and Spohn \cite{PS1} on polynuclear 
growth (PNG), and the creative mathematical efforts of Johansson \cite{KJ} on the single-step (SS) model, both making explicit connections to the related problem of directed polymers in random media (DPRM).  These works established that earlier numerically observed 1+1 KPZ height PDFs were simply zero-mean, unit-variance versions of the universal Tracy-Widom (TW) limit distributions \cite{TW}, well-known from random matrix theory- RMT. 
Here, flat (radial) KPZ growth scenarios corresponding to Gaussian Orthogonal (Unitary) Ensembles; i.e., GOE (GUE), respectively. 
Indeed, TW-GUE was the very same PDF at the heart of the Ulam problem \cite{BDJ,AD99}, a purely mathematical matter focussed on the fluctuation statistics of the longest increasing subsequence (LIS) of a random permutation of  the first $n$ integers. Indeed, at this time, Okounkov \cite{Andrei00} had elegantly established the LIS-RMT equivalence, a connection suggested empirically from the numerical work of Odlyzko and Rains \cite{OR99}, building upon nearly prehistoric efforts of Baer and Brock \cite{BB68}, which indicated the inevitable centrality of TW-GUE limit distribution in both instances.  Pr\"ahofer-Spohn emphasized, as well, the importance of the Baik-Rains (BR) F$_0$ limit distribution \cite{BR}, with no RMT manifestation, dictating temporal correlations of the KPZ stationary-state.  A separate, critical paper by Pr\"ahofer and Spohn \cite{PS2} introduced the notion of the Airy process, bringing  spatial correlations into the RMT context, and tightening ties between KPZ physics and TW mathematical communities.  During this period, flameless Finnish firefronts continued to burn \cite{FFF1,FFF3,FFF5}, mode-coupling \cite{Colaiori} and nonperturbative \cite{HF05,HF06} approaches developed further, 
and a suggestive Dutch chemical vapor deposition experiment appeared \cite{GP}, providing evidence of an {\it asymmetric} 2+1 KPZ height fluctuation PDF.  For a recent example of the nice interplay of KPZ/DPRM statistical physics and TW/BR mathematics, as well as a brief account of  the foundational Ulam-LIS numerics of Baer and Brock, see \cite{Luna}; note, too- \cite{MN,Toom}.  

A 3$^{rd}$ KPZ Revolution- in fact, a veritable KPZ Renaissance \cite{KK,IC}- was inaugurated in 2010 with the elegant, exact, and nearly complete solution of the 1+1 KPZ/DPRM problem. This included the full time-evolution of the universal PDFs to their asymptotic TW forms, managed by independent researchers using complementary driven lattice-gas \cite{SS10,Amir} and replica-theoretic DPRM approaches \cite{CDR,Dotsenko}, in the former instance relying heavily upon recently gained insights into the weakly asymmetric simple exclusion process \cite{TW09}.   The initial advance was for the KPZ {\it wedge} geometry with its TW-GUE connection. Shortly thereafter, the transient dynamics  of the 1+1 {\it flat} KPZ/TW-GOE class was finessed by Calabrese and Le Doussal \cite{Calabrese} using Ising-style Pfaffian tricks within the DPRM context; for DPRM-hard wall interaction, relevant to TW-GSE, see \cite{TG12}. Lastly, the {\it stationary-state} statistics of the 1+1 KPZ equation and its relation to the BR-F$_0$ limit distribution was elucidated via replicas by Imamura and Sasamoto \cite{IS}; later, more rigor \cite{BC}. These extraordinary analytical works were matched, simultaneously, by the experimental efforts of Takeuchi and Sano on turbulent liquid crystal kinetic roughening phenomena.  Reliant upon the machinery of KPZ scaling theory, they first observed TW-GUE statistics \cite{TS10}, then TW-GOE \cite{TS11,TS12} and, finally, identified a key experimental signature of the KPZ stationary-state \cite{KT13,Luna}.  Experimental manifestation of KPZ BR-F$_0$ remains, however, a pressing matter. Importantly, the present  KPZ Renaissance has inspired a renewed interest in the 2+1 KPZ problem, as well as other dimension-dependent issues, such as the existence of a finite UCD, and using Wilsonian renormalization group methods, hopes of elucidating the full KPZ phase diagram \cite{KCW12}.  Universality of the 2+1 KPZ Class and its associated limit distributions (i.e., higher-dimensional analogs of TW-GUE, TW-GOE and BR-F$_0$) have been well-characterized \cite{HH12,HH13,SF13,MP03}, along with universal spatial and temporal correlators \cite{Dutch,Carrasco}, supplemented by secondary KPZ ``patch" PDFs \cite{Dutch,Almeida}.  On the mathematical side, the transmission has shifted into hyperdrive; a rapidly accelerating moveable feast- e.g., \cite{MH,TS,OC,B1,B2,PC14,Hahn,Petrov,DSD15,KJ15}.

\section{An Homage to PS}
\label{sec:1}
In Figure 1, we tip our hat to Pr\"ahofer and Spohn \cite{PS1}, plotting up the flat, stationary, and curved height fluctuation statistics for the {\it KPZ equation  itself,} a counterpoint to the PNG model of the original opus, stacked up against the TW-GOE, BR-F$_0$, and TW-GUE limit distributions. To be clear, in the first instance, our 1+1 KPZ Euler integration, with $\lambda=20,$ uniform stochastic noise, and time step $\delta t=0.01,$ was obtained via a system size $L$=250000, with averaging done over 4000 realizations (thus yielding 10$^9$ data in our statistical sample...), 
and integrated through 2000 time-steps.  To make comparison to TW-GOE, these 1+1 {\it flat} KPZ Class height fluctuations are shifted, rescaled and swapped for the order-1 statistical Tracy-Widom variable $\chi_1$,  according to $\chi_1=(h-v_\infty t)/(\Gamma t)^\beta$, the parameter $\Gamma=\frac{1}{2}A^2\lambda,$ trivially related to KMHH $\theta$ above, and $\beta$=$\frac{1}{3}$ the exact KPZ exponent in this dimension.  With longer runs, suitably averaged, we have extracted the asymptotic growth velocity $v_\infty$=0.195935 for the KPZ equation, as simulated.  Note that, in contrast to PNG, where the phenomenological parameters 
$A, \lambda$ and $v_\infty$ are known exactly, pinning down these quantities represents the primary challenge, beyond simply aggregating the height fluctuation statistics.  In fact, as done in experiment \cite{TS11,TS12} and occasionally in numerics, we have matched the variance of the transient regime statistics (i.e., our 1+1 KPZ Euler pdf) to TW-GOE, fitting the second moment to $\langle\chi_1^2\rangle$=0.63805, which fixes $\Gamma^\beta$=0.23111.  The resulting numerical data set, when viewed against the known TW-GOE limit distribution, focusses attention upon the skewness and kurtosis, as well as the finite-time correction to the universal mean.  In particular, our extracted value for the 1+1 KPZ skewness, $s$=$\langle\delta h^3\rangle_c/\langle\delta h^2\rangle_c^{3/2}$=0.279, as compared with accepted result $s_1$=0.2935. As emphasized recently \cite{Luna}, $\Gamma$ thus determined, can then be employed in the service of a {\it genuine comparison} to both the radial TW-GUE and stationary-state BR-F$_0$
limit distributions.  In other words, the KPZ scaling parameters, while model-dependent, are not subclass dependent- in a given dimension,
the same $\Gamma$ and $v_\infty$ apply to flat, curved and stationary cases, as noted already in experiment \cite{TS12}. Thus, here, we take the numerical short-cut, using TW-GOE to {\it determine} $\Gamma^\beta$, privileging the greater challenges posed by BR-F$_0$ \& TW-GUE. 

Hence, we have also included in Figure 1 the centered, rescaled
fluctuations of the {\it height increment,} $\Delta h=h(t_o+\Delta t)-h(t_o)$, casting the associated {\it stationary-state} KPZ statistics in terms of the $\mathcal{O}(1)$ Baik-Rains variable:
$\chi_0=(\Delta h-v_\infty\Delta t)/(\Gamma\Delta t)^{1/3}$,
\begin{figure}
  \includegraphics[width=0.88\textwidth]{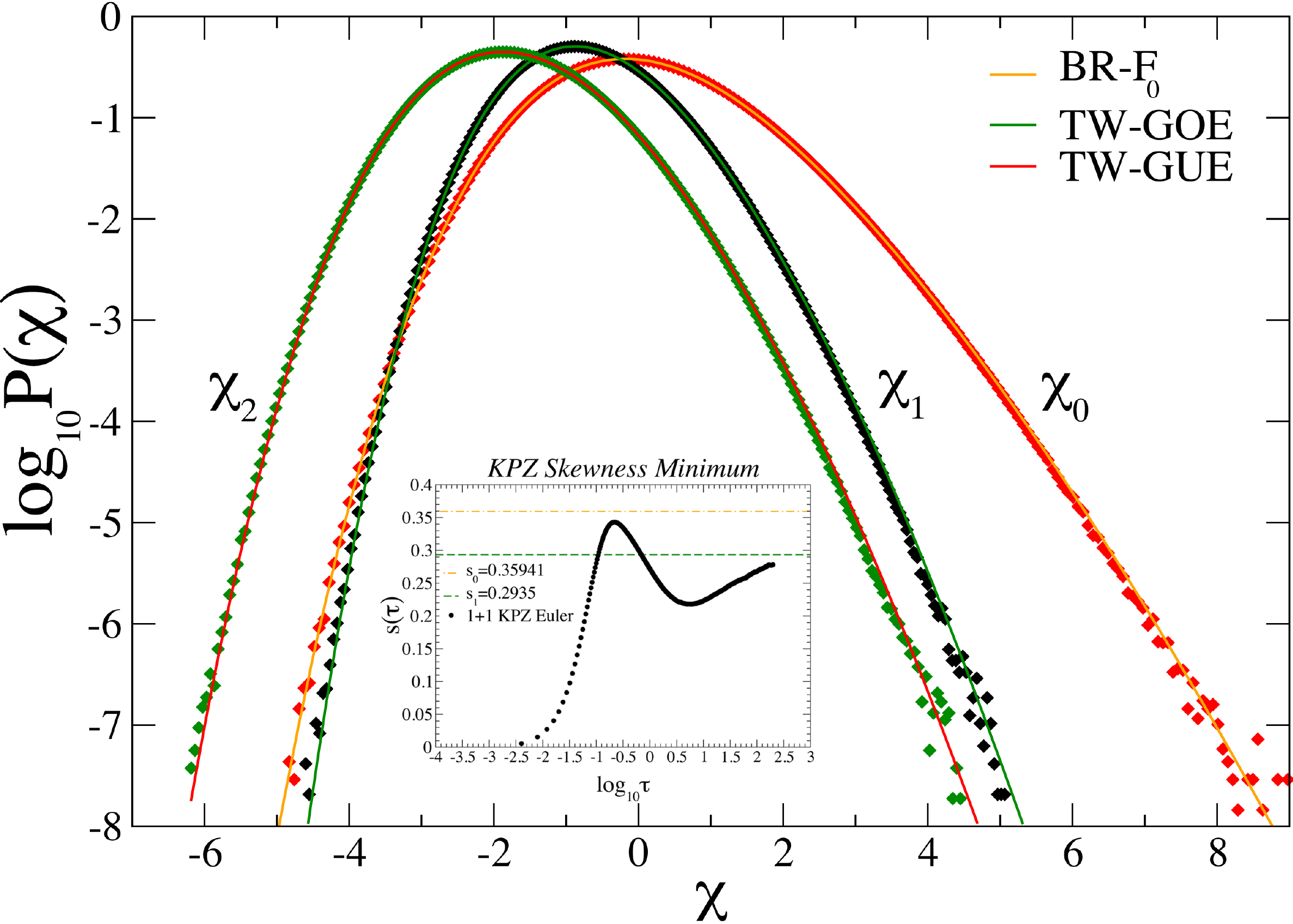}
\caption{1+1 KPZ Euler: Numerical integrations, for {\it stationary-state} ($\chi_0)$, {\it  transient flat} ($\chi_1)$, and {\it radial} ($\chi_2)$ geometries, stacked up against relevant universal limit distributions Baik-Rains, Tracy-Widom GOE, and GUE, respectively.
Inset: Skewness minimum, $s_{TM}\hspace{-0.8mm}\approx$0.22, exhibited by KPZ equation as experimental signature of stationary-state statistics, recently observed in KPZ turbulent liquid-crystal kinetic roughening work \cite{KT13}. Note, in particular, the asymptotic approach from below to TW-GOE skewness $s(\tau\gg$1)=$s_1$=0.2935.        }
\label{fig:1}       
\end{figure}
integrating the KPZ equation to a late time $t_o=5000,$ then investigating the temporal correlations at a 
later time $t_o+\Delta t,$ with $1\le\Delta t\le25\ll t_o,$ at a given site, sampling over the entire system of size $L$=10$^4,$ then averaging over 10$^5$ realizations, again yielding a statistical data set with 10$^9$ points. Here, for the skewness and kurtosis,  we find 0.348 and 0.261, compared to Pr\"ahofer-Spohn's oft-quoted BR-F$_0$ values: $s_0$=0.35941 and $k_0$=0.28916, respectively, for these quantities.
Additionally,  our 1+1 KPZ Euler estimate here for the variance, 1.1527, compares well to the known Baik-Rains constant $\langle\chi_o^2\rangle$=1.15039 and, as is evident from Figure 1, the 1+1 KPZ Euler stationary-state statistics are in fine agreement with the BR-F$_0$ limit distribution, proper.
Note, by construction, this PDF has zero mean; i.e., vanishing first moment.  
As discovered by Takeuchi \cite{KT13}, both experimentally for the 1+1 flat KPZ class turbulent LiqXtal system as well as the PNG model numerically, there is a dip in the distribution skewness as a function of the dimensionless parameter $\tau=\Delta t/t_o,$ as it interpolates between its TW-GOE ($\tau\gg 1$) and BR-F$_0$ ($\tau\ll 1$) values, $s_1$=0.2935 and $s_0$=0.35941, respectively.  The universality of the KPZ Baik-Rains limit distribution, Takeuchi's skewness minimum, and other aspects of the transient to stationary-state crossover for the 1+1 flat KPZ class, were addressed in a detailed, careful study \cite{Luna} examining many diverse models.  In Figure 1 insert, we strike a final note capturing, in a single long run, the skewness minimum in the KPZ equation itself.  
Here, we have $L$=10$^4,$ as before, but now $t_o$=5 and 0.02$\le\Delta t\le$1000, and find $s_{min}$=0.218(4) for $\tau_{min}$=$\frac{\Delta t}{t_o}\hspace{-0.8mm}\approx$5.0, near prior findings \cite{Luna}, which indicated  $s_{min}$=0.225$\pm$0.005.  
We mention, in particular,  that this quantity, being the ratio of moments, requires no knowledge {\it at all} of the KPZ scaling parameter $\Gamma$,
and emphasize its great value as an experimental signature of the KPZ stationary-state.

Lastly, we include within Figure 1, a KPZ Euler integration relevant to the radial fluctuations in the TW-GUE class.  We mention that, aside from some early works \cite{Maritan,trees,Singha}, and more recently \cite{Mas}, there has been little direct effort on numerical integration of 2d KPZ equation in {\it polar coordinates;}
see, too \cite{SS11,SS14,SS15}. Indeed, all work on this subclass, aside from radial Eden model simulations \cite{Alves11,Kaz12}, have resorted to difficult, somewhat frustrating pt-pt simulations of various KPZ/DPRM models \cite{SA13,Luna} in what is, effectively, constrained wedge geometries.
The frustration arises because, in contrast to simulations for the flat KPZ subclass where all substrate points contribute to the ensemble average, the pt-pt Monte Carlo yields a few datum only per realization.   On a practical level, this results in thousand-fold less payoff, given present-day CPU capabilities and equivalent run times; that is, probablilities down to 10$^{-6}$ vs 10$^{-9},$ for
TW-GUE vs. GOE, respectively.  In any case, we report here results for KPZ Euler that rely neither on polar coords, nor a constrained pt-pt Monte Carlo, but rather upon an interesting numerical trick, built upon expanding substrates \cite{Carrasco}.  We refer the reader there for technical details, but the
basic idea is to perform the simulation in the flat geometry, starting with a tiny substrate of system size $L=L_0,$ but increasing the size of the substrate by $\Omega$ sites per unit time, stochastically duplicating column heights of randomly chosen points. This was, effectively, what happened in the polar coordinate KPZ integrations, in any case, since the kinetically-roughened frontier is expanding radially and its discrete mesh must be enhanced; i.e., its population of points grown in time.  In practice, one choses $L_0=\Omega$ so, after an elapsed time $t,$ this expanded substrate has grown to a size $L=\Omega  t.$  In the original work, this method was used to check TW-GUE predictions for three specific stochastic growth models: RSOS, SS and Etching, with typical values $L_0=\Omega=20$ and $t=500$.  A key ingredient involves the addition of a logarithmic time-dependence to the standard KPZ scaling ansatz: $h = v_\infty t +(\Gamma t)^\beta\chi +\zeta {\rm ln}t, $ with $\zeta$ an independent stochastic quantity.  As a warm-up, we have checked this methodology in the directed polymer context, using our well-characterized Gaussian  $g5_1$ DPRM from previous work \cite{Luna}, where we performed a traditional, but quite labor-intensive zero-temperature pt-pt Monte Carlo simulation, with extremal paths of length $t=300,$ averaging over 10$^8$ realizations of the random energy landscape, obtaining $(\langle\chi_2\rangle,\langle\chi_2^2\rangle,s_2,k_2)= (-1.77097,0.8167,0.2304,0.0924),$ quite close to Bornemann's state-of-the-art numbers \cite{Bourne}.  Here, we report an {\it expanding substrates} implementation of this same gaussian polymer model, using $L_0$=20, $t$=2000, and averaging over 25000 realizations, which generates $10^9$ data at the final time slice, resulted in quite similar estimates (-1.77135,0.8311,0.2239,0.09373).  In fact, these new  $g5_1$ DPRM results seem superior to KPZ growth model implementations on expanding substrates; specifically, in the admittedly difficult task of pinning down the kurtosis, which appears systematically underestimated, esp., the Etching model- see Ref\cite{Carrasco}-Fig 2b, 
which indicates $k_2\approx$ 0.07-0.08.  

Performing the same trick for 1+1 flat KPZ Euler, however, presents its own difficulties since long simulation times are not easily accessible.  Nevertheless, with $L_0$=20 and 10$^4$ time-steps, which implies fully expanded substrates of size $L$=2{\rm x}10$^5$, and averaging over 5000 realizations, we obtain here the following 1+1 KPZ Euler estimates for TW-GUE moments: (-1.77682,0.8108,0.2121,0.0826),  
and show, in Figure 1, the associated probability density in $\chi_2$, measured against TW-GUE.  We have taken the height fluctuation PDF at the final time slice ($t$=100), rescaled the width of the distribution with the model-dependent parameter $\Gamma^\beta$=0.23111,
determined precisely by the earlier fit to TW-GOE, and then shifted the curve horizontally to the
asymptotic, but model-specific, mean value $\langle\chi_2\rangle$=-1.77682 dictated by a full, finite-time scaling analysis.  
In any case, the agreement between our new 1+1 KPZ Euler integration and the known TW-GUE trace is reassuring. That said, we mention a slight
residual underestimate of our 2d radial KPZ kurtosis $k_2$, seen previously for  the deposition growth models.

\section{ d=1+1 KPZ Class Experiments: Stationary-State Metrics via WBC}
\subsection{KPZ ``Patch"-PDF I: Roughness Distributions for 1/$f^2$ Noise \& Wiener Paths}
Beyond classic Tracy-Widom and Baik-Rains limit distributions capturing universal properties of kinetically roughened edges in the 1+1 KPZ Class, there exist additional PDFs of relevance which characterize various intriguing aspects of the KPZ {\it stationary-state statistics.} We discuss in the following two such distributions which, experimentally and numerically, can be accessed via locally-averaged, so-called ``patch" PDFs tied to sampling at small scales.  These PDFs can play an important supporting role in the analysis of stochastic growth phenomena and have been conceived
with the exigencies of the KPZ experimentalist well in mind, privileging window boundary conditions (WBC) over the theorist's favorite- PBC. For an early pre-TW/BR indication of these practicalities, dating to the 1$^{st}$ KPZ Revolution, we mention the efforts of R\'acz and Plischke \cite{Racz}.
A longstanding statistical quantity of interest has been the squared-width: $w^2$=$\langle h^2\rangle$-$\langle h\rangle^2$;
here, angular brackets signifying a spatial average over a tiny microscopic patch of linear dimension
$\ell$, smaller than the finite dynamical correlation length $\xi_\parallel.$ With an ensemble average over snapshots in the experimental system, the task is to aggregate the statistics of $\omega=(w^2-\langle w^2\rangle)/\sigma^2_{w^2}$, which is the centered, rescaled, dimensionless order-one fluctuating statistical variable at the core of the universal  {\it roughness distribution-} $P_{RD}(\omega ).$  Recently, this metric was applied in 2+1 KPZ Class vapor deposition experiments involving organic \cite{Dutch}  and semiconductor \cite{Almeida} thin films.  In the former instance, Halpin-Healy and Palasantzas compared experimental data directly to their numerical 2+1 KPZ Euler RD, using WBC ``box"-sizes $\ell\ll\hspace{-1.5mm}\xi_\parallel^{KPZ},$ for which the statistics were {\it stationary,}  yielding nearly constant quantities. The 2+1 KPZ Euler RD was characterized by a skewness $s$=2.03 and kurtosis $k$=7.11, while the organic thin-film experiment yielded measured values: $(s,k)$=(2.12,7.77). Since the roughness distribution, by construction, is a zero-mean, unit-variance PDF, $s$ \& $k$ become the two essential quantities to pin down. 
Thus, the agreement is quite suggestive and given the great demands of thin-film experiment, as well as the relative paucity of temporal data there, reveals $P_{RD}(\omega)$ to be a welcome, supplementary measure \cite{Racz} of 2+1 KPZ kinetic roughening.  

\begin{figure}
\includegraphics[width=0.99\textwidth]{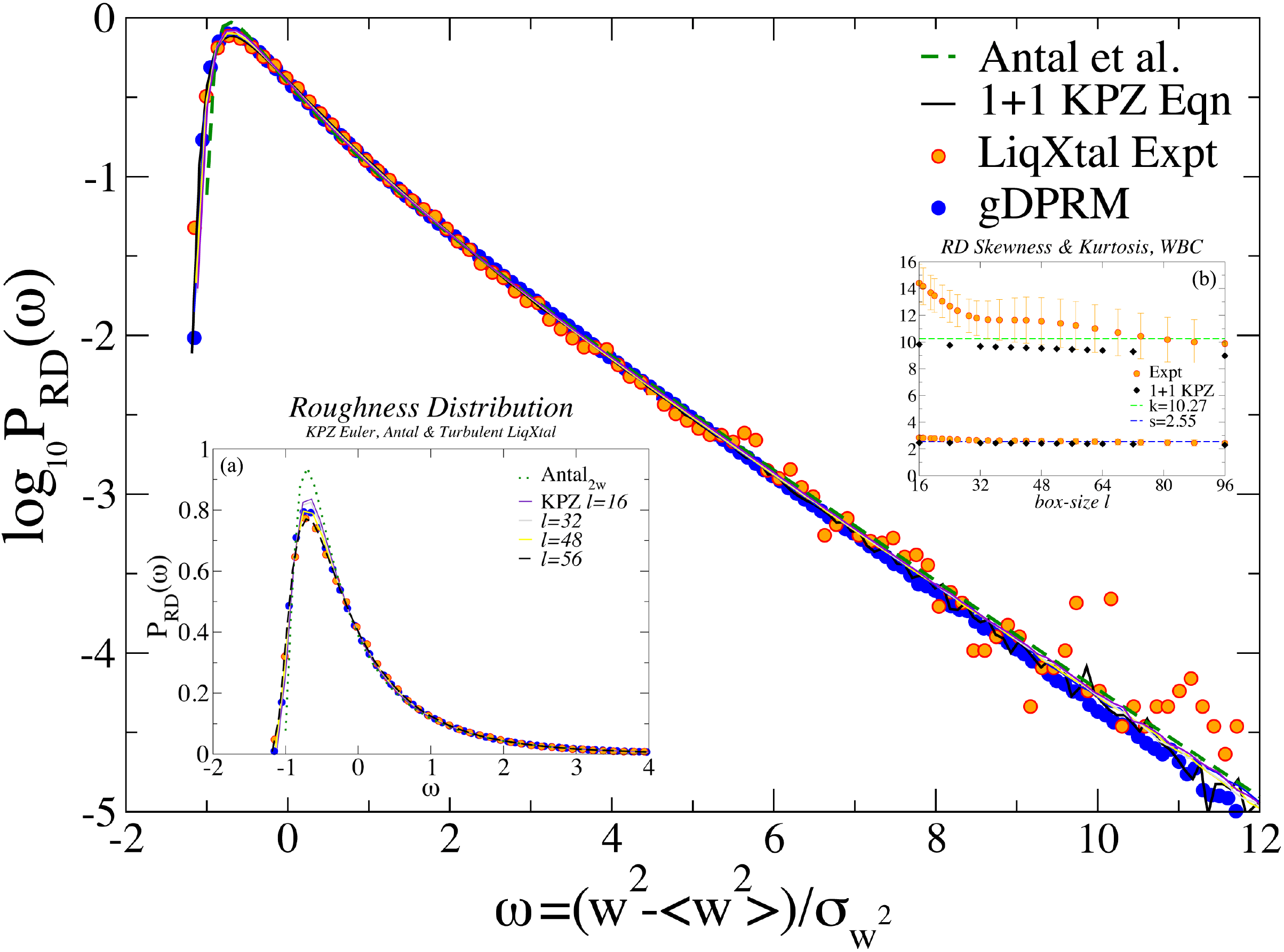}
\caption{Experiment, Numerics \& Theory: 1+1 KPZ Class RD, semilog plot. Comparison- 1+1 KPZ Euler numerical integration, Gaussian 
DPRM, turbulent liquid-crystal data \cite{TS10,TS11,TS12} and the roughness distribution of Antal {\it et al.} for 1/f$^2$ noise \cite{Antal}. Insets: a) Linear plot. Near the peak, thin KPZ Euler traces move upwards for decreasing box-size. b) RD skewness and kurtosis, as a function of patch-size $\ell.$  Dashed horizontal lines indicate associated Antal$_{2w}$ values.  In experimental system, $\ell$ is measured in pixels, corresponding to 4.74 microns. }
\label{fig:1}
\end{figure}

Motivated by these 2+1 KPZ vapor deposition experiments, we have revisited the 1+1 KPZ Class turbulent liquid-crystal work of Takeuchi and Sano \cite{TS11,TS12}, examining the data from the perspective of patch statistics and the local roughness distributions.  Conveniently, our efforts here have been preceded by two extremely helpful theoretical papers. The first, by Foltin {\it et al.} \cite{Foltin}, calculates the {\it exact} roughness distribution of 1d {\it random-walk}  Edwards-Wilkinson (EW) interfaces \cite{EW},  assuming PBC. We remind the reader that d=1+1 is a very special case and, as regards the steady-state statistics, the KPZ nonlinearity is irrelevant in this dimension, with both EW 
and nonlinear KPZ proper {\it sharing the same stationary-state measure.} While this statement is typically voiced with the thermodynamic limit in mind, it is also true for a finite system with PBC; see Fokker-Planck discussion, next section. In fact, these authors numerically verify the 1+1 EW/KPZ steady-state equivalence in this PBC context, extracting the roughness distribution of catalytic reaction model, honorary member of the 1+1 KPZ Class; see \cite{Foltin}-Fig 4 there, which reveals very fine KPZ agreement (in fact, better than EW!) with their exact RD for periodic random walks.  That's great, but we consider patch-PDFs using experimentally-dictated WBC. Serendipitously, later work by Antal and coworkers \cite{Antal} addresses precisely this case; i.e., finite Brownian paths with WBC. Truth be told, their paper considers the much more general issue of RDs of periodic and nonperiodic 
$1/f^a$ noise. Interestingly, their results establish that, generically, the respective RD PDFs are distinct, PBC vs. WBC- see \cite{Antal}-Fig 6, which makes this quite clear. Here, $a$=2 corresponds to 1+1 random-walk EW interfaces, and they obtain the exact roughness distribution for WBC;
see Ref.\cite{Antal}-sect. IV.A-B \& Figs. 4 \& 5, which illustrates the relevant Brownian RDs.  In Figure 2 above, we compare this Antal$_{2w}$ RD with Takeuchi-Sano liquid-crystal data sets, as well as our newly-obtained 1+1 KPZ Euler WBC roughness distribution- here, $\lambda$=20, with system size $L$=10$^5$, and averaging done over 1000 runs.  The main plot is semi-logarithmic,  fully revealing the long tail, with probabilities down to 10$^{-5}$; inset left-Fig. 2a, illustrates the straight-up RD with linear scales.  Additionally,  statistics for a Gaussian polymer model, $g5_1$ DPRM discussed previously, with normally distributed site energies has been included to indicate universality of our 1+1 KPZ/DPRM results. 
Finally, Fig. 2b records the $\ell$-dependence of the 1+1 KPZ RD skewness and kurtosis which, with diminishing box-size $\ell$ \cite{Santa}, is observed to approach
values characteristic of the Antal$_{2w}$ RD, calculated by us to be $(s,k)$=(2.55,10.27), both $s$ and $k$ distinctly larger than 2+1 KPZ RD counterparts \cite{Dutch}: (2.03,7.11).  
On the experimental side, the LiqXtal RD skewness rides slightly above 1+1 KPZ, but the kurtosis suffers a bit more from the noisy statistics at the end of that long right-tail, which we extend nearly 12 standard deviations. We mention, as an aside, digging that far into the right tail of Antal$_{2w}$ RD trace demanded five terms in the Laplace transform series, rather than the three of the original paper \cite{Antal}.  In this regard, we see pure {\it exponential} behavior for WBC, as is already known for periodic boundary conditions, though with a smaller prefactor than the $\pi^2/6$ characteristic of Antal$_{2p},$ appropriate to PBC \cite{Foltin}.  All this, of course, contrasts with the 2+1 KPZ Euler roughness distribution  \cite{Dutch} which, for WBC, appears to possess a stretched exponential tail with exponent 7/8.

\subsection{{\bf KPZ ``Patch"-PDF II:  Extremal Statistics \& Majumdar-Comtet Distribution}}
Our second patch-PDF focusses on the extreme-value (EV) statistics of the height fluctuations, providing a kinetic roughening context in which to study extremal behaviors of {\it correlated} random variables. In this setting, a natural statistical variable is the maximum-relative-height (MRH), $m\equiv h_{max}-\overline{h},$ measured with respect to the mean interface position in the patch.  Extremal MRH fluctuations on the global scale, in large systems assuming PBC, was first examined numerically \cite{Shapir} for the Edwards-Wilkinson case, but then solved exactly in 1+1 dimensions in beautiful work by Majumdar and Comtet \cite{Comtet,Comtet2}.  Somewhat surprisingly, they discovered that the Airy PDF, famously implicated in dictating areal statistics beneath Brownian bridges, was at work here as well, controlling MRH fluctuations of 1d {\it random-walk} EW interfaces.    As these authors emphasize, and check, the Airy distribution also governs MRH statistics for the 1+1 KPZ stationary-state \cite{Comtet,Comtet2,GS}. This is not unexpected, since the KPZ nonlinearity generates boundary terms which vanish, yes, in the asymptotic limit, but also in finite systems with PBC; in other words, for PBC, the same stationary measure solves the Fokker-Planck equation for both 1+1 EW and KPZ, a well-known special feature unique to this dimensionality.   Unfortunately, as pointed out by Majumdar and Comtet, this bit of magic does {\it not} occur for the WBC relevant to us.  Additionally, they calculate a quite distinct PDF relevant for 1+1 EW interfaces with ``free" (i.e., Neumann)  BC, which they christen ``F-Airy"; see, Ref.\cite{Comtet}-Fig 1, for a comparison of these two universal PDFs.
 
Given the missing magic,  as well as the additional potential wrinkle of Neumann versus strict window boundary conditions, there is no reason necessarily to expect that this new F-Airy distribution will characterize the 1+1 KPZ Class WBC situation too- for key details, we refer to Ref.\cite{Comtet2}-sect. 5.  Interestingly, despite these caveats, our analysis suggests that the Majumdar-Comtet PDF does, in fact, capture the stationary-state MRH fluctuations of 1+1 KPZ Class, thereby preserving the KPZ-EW steady-state equivalence and 1d random-walk interpretation, WBC notwithstanding.  We direct the reader's attention to Figure 3 where, employing WBC, we examine MRH statistics in local patches of size $\ell\ll\hspace{-1.5mm}\xi_\parallel^{KPZ}$ showing turbulent LiqXtal data sets, and making explicit comparison to 1+1 KPZ Euler integration and $g5_1$ DPRM simulations analyzed from this perspective. 
Again, the main point is that by restricting ourselves to sufficiently small $\ell$ values, we are actually accessing the 1+1 KPZ stationary-state.
We include, too, the Majumdar-Comtet F-Airy distribution, proper. All this is presented in a zero-mean, unit-variance rendering to highlight the skewness and kurtosis of these PDFs. The main plot is semi-logarithmic, revealing the F-Airy tails in their full glory; for large, positive $\mu=(m-\langle m\rangle)/\sigma_m$, the behavior is known \cite{Comtet} to be {\it quadratic.} That is, a gaussian tail, seen in the central inset, which uses linear scales. A second inset, upper right, reveals the $\ell$-dependence of the skewness and kurtosis of our 1+1 KPZ Euler MRH patch-PDFs. Here, dashed lines correspond to the values $(s,k)$=(1.11,1.69) characteristic of the Majumdar-Comtet distribution, which we have extracted via its 10-term confluent hypergeometric representation \cite{Comtet2}. Our thick KPZ Euler curve corresponds to $\ell$=56, where we have $(s,k)$=(1.01,1.48).  As the patch-size shrinks, however, greater stationarity is insured and agreement improves- see thin traces, corresponding to $\ell$=48,32,16; this is especially evident in the center inset, near the peak of the F-Airy distribution. Hence, for $\ell$=32, our KPZ Euler MRH patch-PDF possesses $(s,k)$=(1.05,1.57), which increases further to (1.07,1.62) for $\ell$=16, clearly heading towards calculated F-Airy values.  We find these vanishing differences rather suggestive and note, in passing, that the Airy PDF appropriate to 1+1 KPZ/EW with {\it periodic} boundary conditions \cite{Gyorgyi2,Rambeau2} has $(s,k)$=(0.701,0.560), certainly quite distinct from Majumdar-Comtet F-Airy, relevant for WBC, which we illustrate in Figure 3 and see in the LiqXtal experiment.

\begin{figure}
\includegraphics[width=0.99\textwidth]{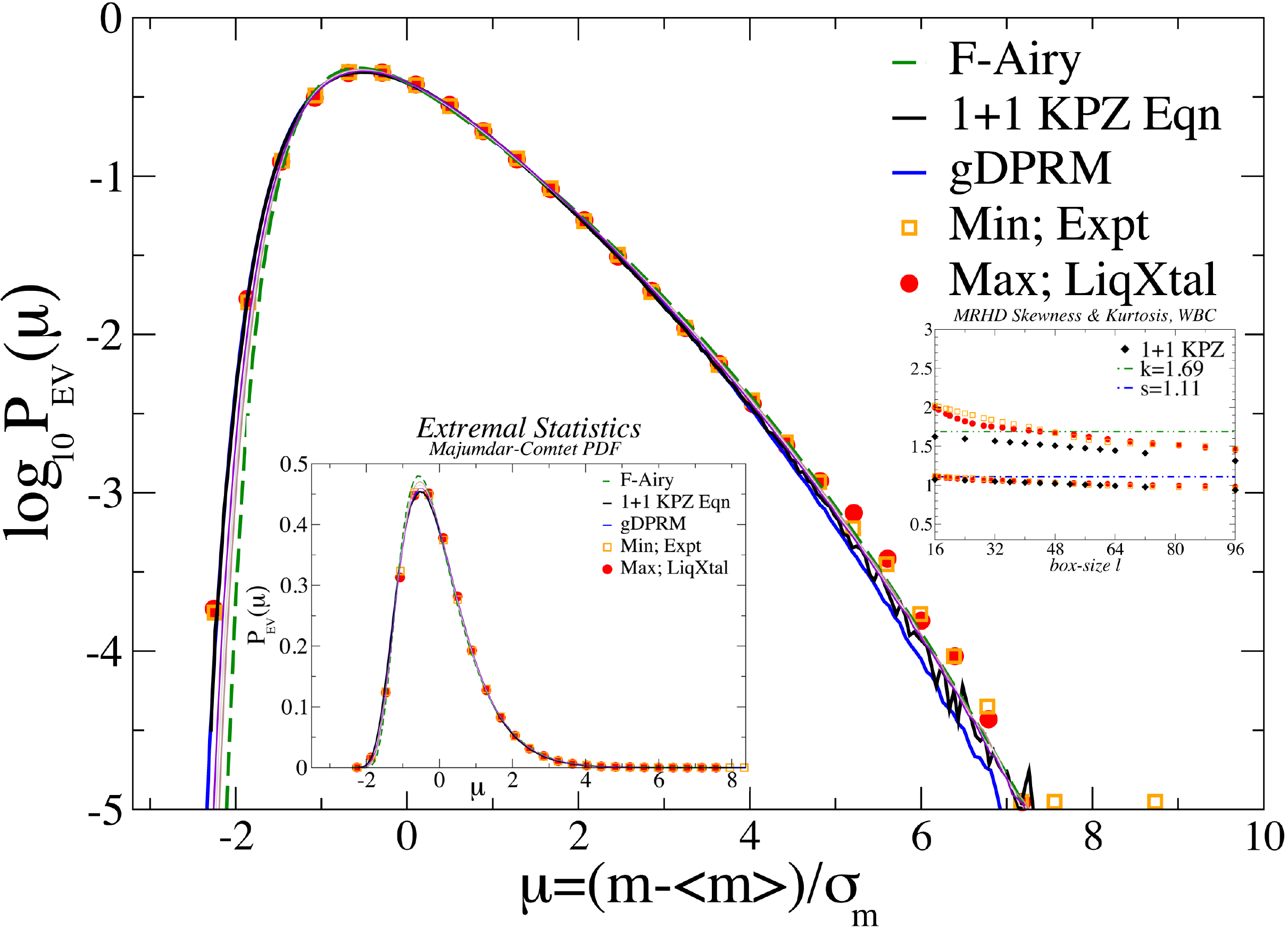}
\caption{1+1 KPZ {\it Stationary-State:} Extremal Height Fluctuations vs. Majumdar-Comtet F-Airy Distribution.}
\label{fig:1} 
\end{figure}

Of course, the MRH distribution governing stationary, extremal height fluctuations of 2+1 KPZ {\it surfaces} presents a substantial theoretical challenge, as  there are no ties {\it at all} to an EW counterpart \cite{DSL} in this higher dimension, regardless of BCs.  Nevertheless, there has been recent progress on 2+1 KPZ Class extremal-height numerics, where explicit comparison has been made with MRH  patch-PDFs obtained from thin-film stochastic growth experiments \cite{Dutch,Almeida}.  An interesting question there then concerns the higher-dimensional analog of the 1+1 KPZ Class {\it gaussian} F-Airy tail; at first glance, the 2+1 KPZ Euler MRHD \cite{Dutch} with $(s,k)$=(0.884,1.20), though surely different, bears some familial resemblance to the Majumdar-Comtet distribution.  Even so, it would be rather unexpected were the 2+1 KPZ MRH right-tail asymptotics also dictated by a precise gaussian fall-off.  In any case, this would be quite a difficult matter to resolve via numerics.

\section{Universal Limit Distribution: 3d Radial KPZ Class}
In previous works  \cite{HH12,HH13}, attention was focussed on large-scale numerical simulations of 
multiple models (RSOS, EDEN, driven-dimer, DPRM), as well as a direct Euler integration of 2+1 KPZ equation itself,
establishing universality of the 2+1 {\it flat} KPZ Class, manifest in a limit distribution $P_{KPZ}^{2+1}(\xi)$ possessing universal moments   
$(\langle\xi_1\rangle,\langle\xi_1^2\rangle, s,k)$= (-0.85,0.235,0.424,0.346), with $\xi=(h-v_\infty t)/(\theta t)^\beta$. 
We draw particular attention to
\cite{HH13}-Section III, Figs. 6 \&7, where a refined portrait of this characteristic PDF, higher-dimensional analog of TW-GOE, 
is presented.  See, too, \cite{HH12}-Table I \& Fig 2 there, where an exhaustive analysis, employing KPZ scaling theory \cite{KMHH}, built with the KM toolbox, permitted determination of all relevant model-dependent parameters.  In particular, for the highly nonlinear, $\lambda=20$, 2+1 KPZ Euler integration, the asymptotic growth velocity $v_\infty=0.17606$, and key scaling parameter $\theta=A^{1/\alpha}\lambda=1.192{\rm x}10^{-3}$.  Indeed, with knowledge of these parameters, it was also possible to isolate the limit distribution dictating the late-time 2+1 KPZ {\it stationary-state} statistics, analog of 1+1 KPZ Class BR-F$_0$. This was done for 2+1 KPZ Euler, gDPRM, and RSOS stochastic growth models, demanding quite long integration times, but resulted in a convincing, composite portrait.  The KPZ stationary-state is quite important as a subclass because it represents the natural contact point with field-theoretic, renormalization group and mode-coupling analyses; here, Halpin-Healy extracted the universal variance of the 2+1 KPZ stationary-state statistics, obtaining $\langle\xi_o^2\rangle=0.464$, analog of the 1+1 Baik-Rains constant $\langle\chi_o^2\rangle=1.15039$.  When properly translated \cite{HH13} into the context of a recent field-theoretic, Wilsonian RG calculation of Kloss, Canet and Wschebor \cite{KCW12}, this becomes the universal renormalization group amplitude $R_{2+1}=0.944\pm0.031$, in fine agreement with the KCW value 0.940. Hence, we see that  $\langle\xi_o^2\rangle=0.464$, where field-theoretic RG meets Monte-Carlo numerics has become, suddenly, the best measured quantity of the 2+1 KPZ problem.  We remain hopeful that the field theorists will someday succeed in crafting their own renormalization-group portrait of the entire limit distribution, or perhaps the skewness therein, known to be $s_0=0.244;$ see Ref \cite{HH13}, Sect. IV and Fig. 9 in that paper.  Interestingly, the 2+1 KPZ stationary-state has also permitted a solid {\it multi-model}
(gDPRM,RSOS,KPZ Euler) estimate $\beta_{2+1}$=0.241(1), in line with prior gold-medal studies \cite{FT,KO} at the 3-digit precision level. 

\begin{figure}
\includegraphics[width=0.99\textwidth]{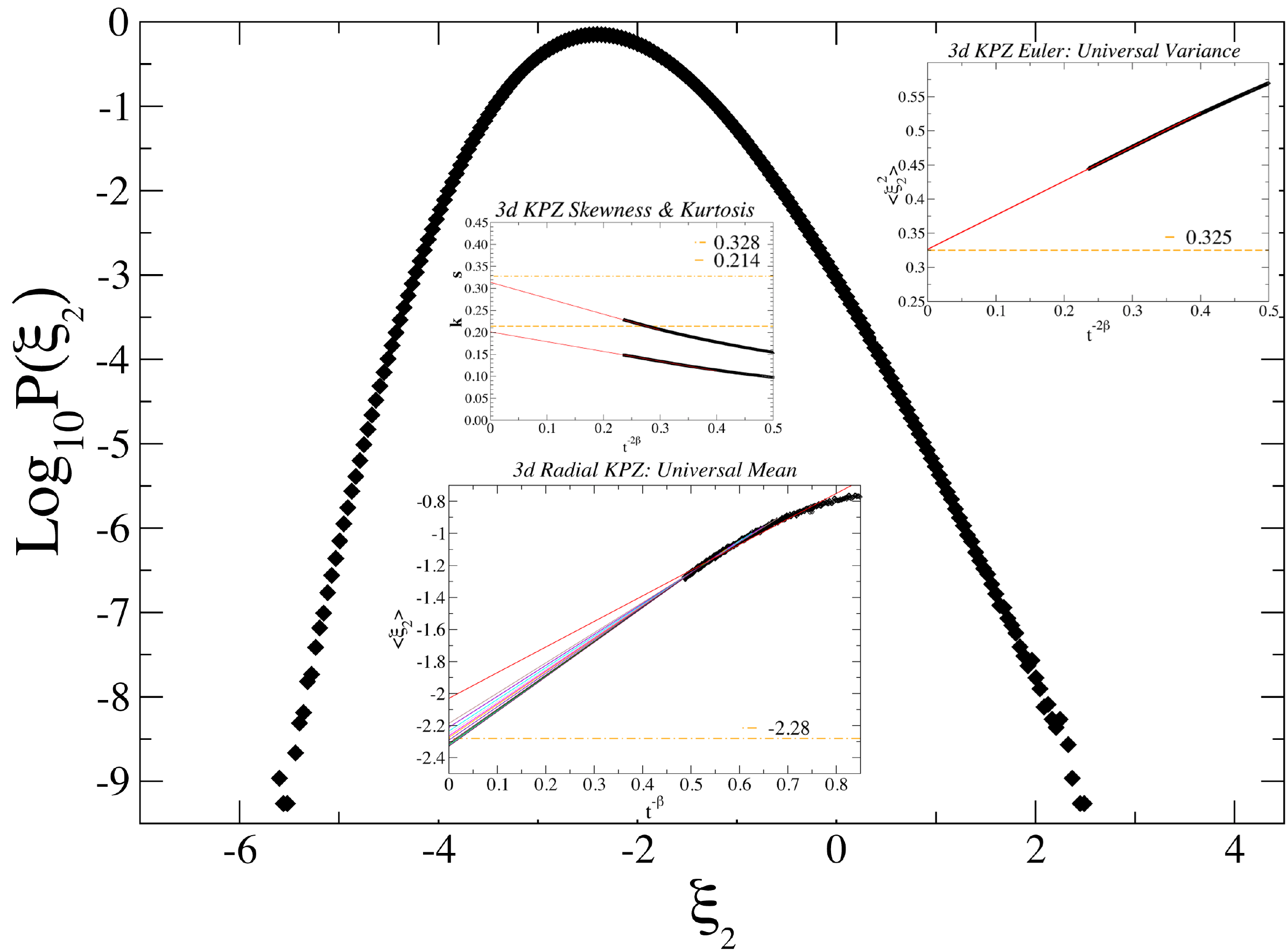}
\caption{Universal Limit Distribution: 3d {\it radial} KPZ Class, via expanding substrates. Insets reveal
finite-time scaling analyses necessary to elicit asymptotic estimates for universal moments. Dashed horizontal lines indicate
known values extracted from independent 3d pt-pt DPRM, SHE, and Eden simulations \cite{HH12,HH13}.}
\label{fig:1} 
\end{figure}

A more stubborn challenge involves unearthing the higher-dimensional KPZ analog of TW-GUE; i.e., the limit distribution characteristic of the 3d {\it radial} KPZ problem.  The 2010 tour-de-force exact solutions notwithstanding, solving the KPZ equation {\it in 3d spherical coordinates,} or the related 3d pt-pt stochastic heat equation (SHE) is a humbling prospect, indeed.
Fortunately, recent numerics have helped chart the course, framing the essential features and providing a target for future analytical efforts.  Halpin-Healy
\cite{HH12} simulated 3d extremal paths connecting far corners of a cube filled with exponentially distributed site energies; the underlying pt-pt energy fluctuation limit distribution was measured to possess universal variance, skewness and kurtosis: 0.291, 0.331, and 0.212, respectively. 
Later authors \cite{SF13}, studying three distinct 3d pt-pt KPZ growth models, among them single-step (SSC) and on-lattice version of Takeuchi Eden D, find values in the range $s_2\approx$ 0.32-0.34 and $k_2\approx$ 0.20-0.22, providing additional evidence for these characteristic, universal quantities. Tucked away in his doctoral dissertation \cite{MP03}- Table 7.2, Pr\"ahofer presciently records $s_2$=0.323(5) and $k_2$=0.21(4) for his own early 3d pt-pt PNG simulation. 
Subsequent work \cite{HH13}, see particularly sect. II: Table I and Fig. 4 there,
which involved direct numerical integrations of the constrained 3d pt-pt SHE, as well as three additional 3d pt-pt DPRM models, has established 
$(\langle\xi_2\rangle,\langle\xi_2^2\rangle, s_2,k_2)$= (-2.28,0.325,0.328,0.214) in this dimension.
Here,  we perform a 2+1 KPZ Euler integration on expanding substrates with $\Omega=10$, complementing our earlier efforts on the 3d pt-pt SHE with multiplicative noise. Results are indicated here in Figure 4, with insets demonstrating that the variance, found to be 0.326, is dead-on, the skewness and kurtosis well within accepted values \cite{HH12,HH13,SF13,Carrasco}, while subleading corrections render extraction of the universal mean a bit more challenging.   Nevertheless, successive fit lines monotonically approach a well-defined boundary (i.e., envelope), with intercepts converging to $\langle\xi_2\rangle\approx$-2.32, a quite decent value.

\section{Diamonds in the Rough: The Hierarchical $\diamondsuit$DPRM Revisited}
\subsection{{\bf From Gauss to Gumbel \& back again}}

The dimensional behavior of the {\it Euclidean} KPZ/DPRM problem can be found, in nearly its full glory,
within a directed polymer implementation on $b-$branched, diamond {\it hierarchical} $(\diamondsuit)$ lattices; see Figure 5, lower left, for geometric reminder. At the heart of the $\diamondsuit$DPRM one finds a simple recursion relation \cite{DG89,HH89}, whose fixed point function, see Fig. 5 for examples, is the much-desired, sought-after limit distribution:
$$\int_\xi^\infty P_{n+1} = \bigg[\int_\xi^\infty P_n*P_n \bigg]^b$$
\noindent Here, $P_n$ represents the  $\diamondsuit$DPRM bond-energy PDF at the $n^{th}$ generation, and  the asterix indicates convolution. The pioneering effort in this regard,
dating to the 1$^{st}$ KPZ Revolution, is due to Derrida and Griffiths \cite{DG89}, an elegant work which quickly inspired an analysis of its many-dimensional generalization \cite{HH89}, addressing the question of a finite UCD in the KPZ context.  Indeed, it was upon these lattices that universal DPRM PDFs were first examined- see, esp., Halpin-Healy \cite{HH90}-Fig. 8 there, which shows the asymptotic evolution of underlying distributions, as well as
\begin{figure}
\includegraphics[width=0.99\textwidth]{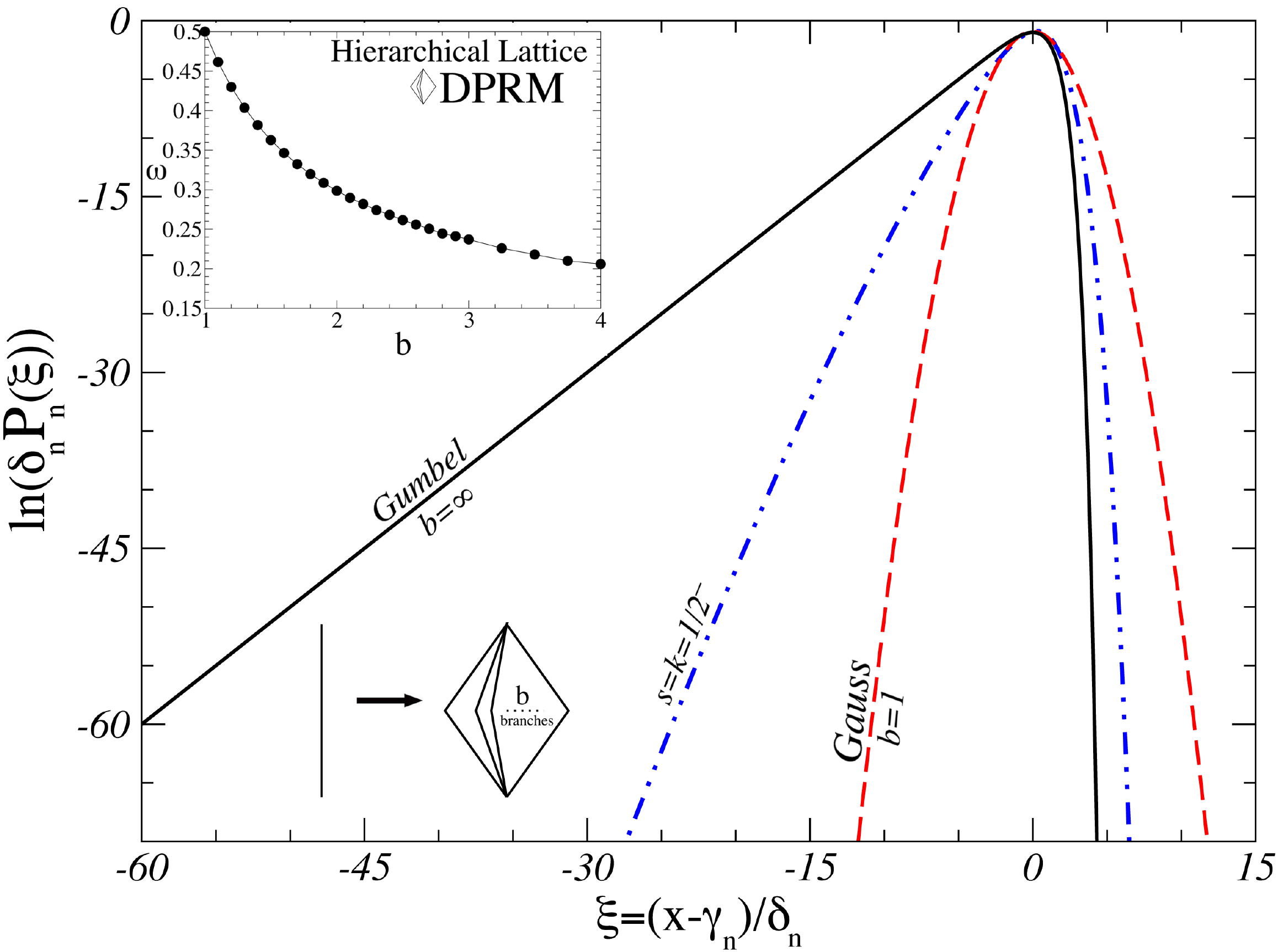}
\caption{Universal Limit Distributions: $\diamondsuit$DPRM Class-  Lattice branching parameters $b$=1, 2.82826, and $\infty$, the first and last values yielding normal {\it Gaussian} and standard {\it Gumbel} PDFs, respectively.  Inset, upper left: $b$-dependence of the KPZ scaling exponent $\omega;$ lower left: Geometric gnomon, $b$-branched hierarchical lattice. }
\label{fig:dDPRM}       
\end{figure}
Derrida \cite{BD90}- Fig. 3, providing a representative snapshot of a, now, very familiar (or, at least very familiar-{\it looking....}) asymmetric curve.
Following up on the $\diamondsuit$DPRM, Roux {\it et al.} \cite{Roux91} proposed an index relation $\eta'=2\eta$, connecting exponents dictating the stretched exponential tails of the universal DPRM energy PDF.  With Zhang's saddle-point formula for the left-tail exponent, $\eta$=1/(1-$\omega)$ and the known 1+1 KPZ/DPRM result $\beta_{KPZ}=\omega_{DPRM}=\frac{1}{3}$ in hand, this led to the conclusion,
addressed numerically \cite{KBM} back in 1991, that $\eta=\frac{3}{2}$ for the long, TW Airy tail, but also predicted $\eta'$=3 for the short, computationally intractable, steep opposing tail. These values, of course, were understood later by Pr\"ahofer and Spohn,  as characteristic of the underlying universal Tracy-Widom limit distributions. Meanwhile, the $\diamondsuit$DPRM lay dormant [indeed, for some time....] until Monthus and Garel (MG) returned  to this rich, deceptively simple toy model.  In an intriguing paper  \cite{MG08}, they 
i) announced a {\it revised} tail exponent conjecture: $$\eta'=d_{\rm eff}\eta$$ for the DPRM problem, ii) provided a saddle-point analysis tailored to and specifically valid for the $\diamondsuit$DPRM, where $d_{\rm eff}=1+{\rm log}_2b$ serves as the {\it fractal} dimension for $b$-branched hierarchical lattices, but also, iii) proposed a simple back-of-the-envelope argument extending the result to  the standard Euclidean DPRM, where $d_{\rm eff}=d+1,$ literally. Lastly, the curious researcher might also compare MG-Fig 3a to \cite{KBM}-Fig 9; the similarity is not incidental, but rather intentional and, indeed, quite germane to the discussion that follows.

As for  the Monthus-Garel conjecture, we direct the dedicated reader to Sections 2, 5.1, as well as Appendix A of this interesting work \cite{MG08}, which contrasts their own rare events analysis to the earlier argument
of Zhang \cite{HHZ} that had first fixed the TW Airy tail exponent $\eta$=1/(1-$\omega$). We  also draw specific attention to MG's numerical confirmation of Zhang's formula for the hierarchical $\diamondsuit$DPRM; esp., MG-Table 1, where they explore numerous integer values of the branching parameter, running from $b$=2, where $\omega$=0.299, so $\eta$=0.143, to $b$=24, where $\omega$=0.123 yields $\eta$=1.14. Unfortunately, the more challenging task of actually testing the MG conjecture, itself, was sabotaged by the large corrections to scaling attendant to $\eta'.$ Nevertheless, the MG data appear ``compatible" with their suggested tail exponent identity.  Finally, the authors present an impressive group portrait of the associated limit distributions for $b$ in this range; see, MG-Fig. 3b.

We have revisited the $b$-dependence of the $\diamondsuit$DPRM universal PDF and, show in Figure 5, our own portrait focussing on three {\it seminal values} of $b$.  That is, the limit $b\rightarrow1$, i.e., $d$=0, the model degenerates to simple sums of uncorrelated random variables, the variance grows linearly, $\omega$=1/2, and the CLT demands {\it symmetric Gaussian;} in other words,  
$\eta$=$\eta'$=2, and $s$=$k$=0.  Conversely, as $b\rightarrow\infty$, where there are an infinite number of branches, one effectively selects the extremum of an asymptotically large set of iid's. Indeed, in this limit, one retrieves the textbook, extremal statistics {\it asymmetric Gumbel} distribution
\cite{Gumbel}, which possesses a skewness $s_G$=12$\surd{6}\zeta(3)/\pi^3,$ involving Ap\'ery's constant, an excess kurtosis $k_G$=$\frac{12}{5},$ a simple exponential for the long, left-tail, and a deep-diving, double exponential for the steep right-tail. Thus, the $\diamondsuit$DPRM provides a convenient {\it continuous} model interpolating between normal Gaussian ($e^{-\xi^2}$) and standard Gumbel ($e^{\xi-e^\xi}$) limit distribution behaviors for $b\in[1,\infty)$, as we emphasize in the figure. The key point here is that, as $b$ grows, the long left-tail swings up and $\diamondsuit$DPRM exponent $\eta$ drops from 2 to 1, evolving from quadratic to linear, while the index $\eta'$ starts at 2 but then diverges without bound, transforming the steep right-tail into a double exponential $e^{-e^\xi}$ in the infinite-$b$ limit.
A final inset to Figure 5 reveals the $b$-dependence of the characteristic KPZ scaling exponent $\omega(b)$, which controls the growth, at the $n^{th}$-generation of the lattice, of the $\diamondsuit$DPRM PDF width $\delta_n\sim 2^{n\omega}$, as well as the vanishing deviation of the distribution mean $\gamma_n$, from its asymptotic value; the latter behavior not shown but, interestingly, {\it fully consistent} with Euclidean KPZ expectations \cite{FF}. 

\begin{figure}
\includegraphics[width=0.99\textwidth]{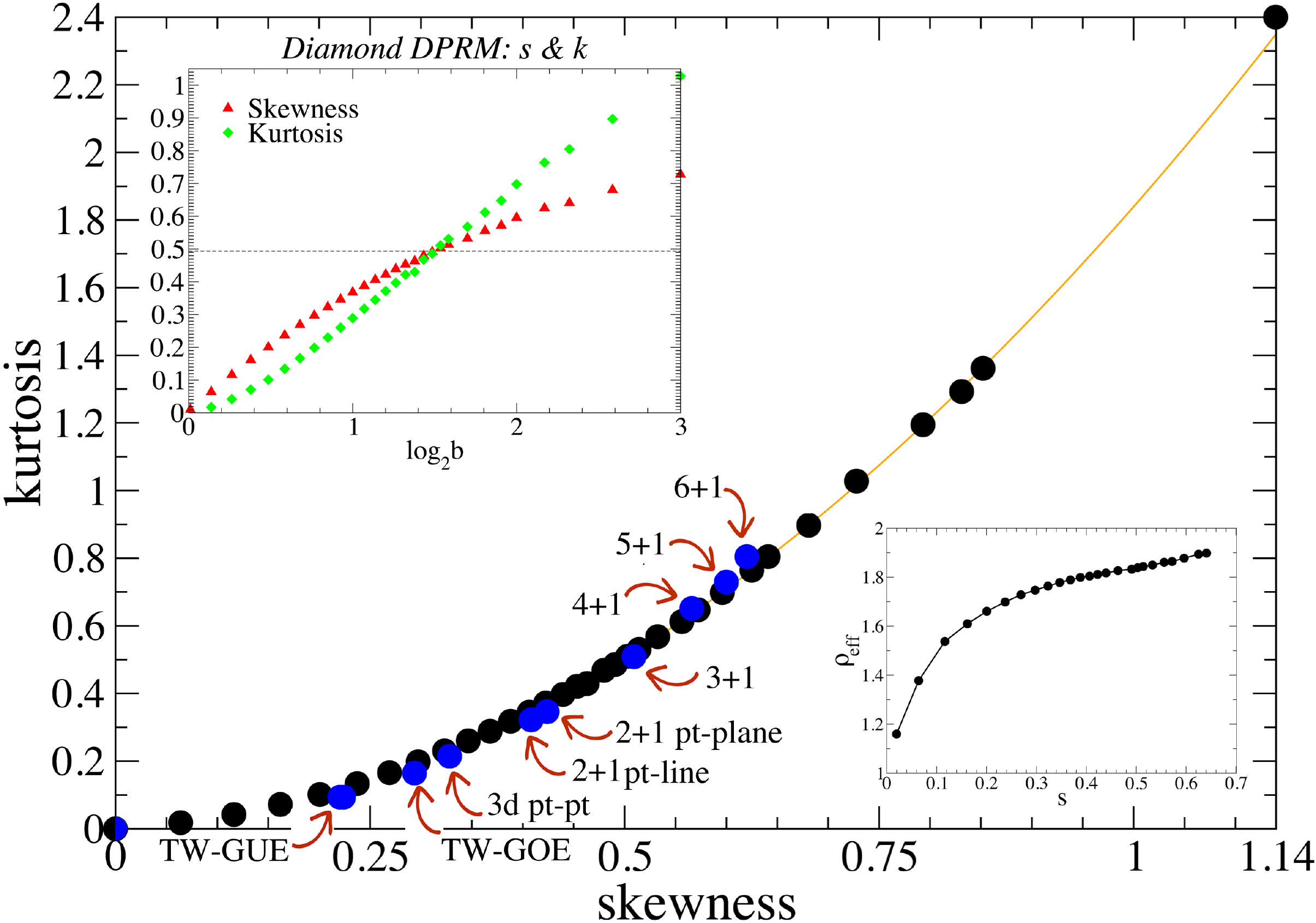}
\caption{Hierarchical ($\bullet$) vs. Euclidean (${\color{blue}\bullet}$)  DPRM: $sk$-plot. Power-law fit to the final dozen points of the $\diamondsuit$DPRM data set, indicated by the colored trace ({\color{orange}--}), nearly pierces the bull's-eye at the upper right corner of the plot frame, its expected destination, where reside the standard Gumbel values $(s_G,k_G)$=$(\frac{12\surd{6}}{\pi^3}\zeta(3),\frac{12}{5}).$ Inset- Upper left: $b-$dependence of the $\diamondsuit$DPRM skewness and kurtosis. They cross, for $b\hspace{-0.8mm}\approx$2.82826, at the common value $s$=$k\hspace{-0.8mm}\approx$0.49375, indicated by the dashed line.   Lower right: Running exponent $\rho_{\rm eff}$ in the $\diamondsuit$DPRM power-law $ks^\rho$-relation. }
\label{fig:1}       
\end{figure}

Here, we concentrate specifically on the $\diamondsuit$DPRM skewness and kurtosis; see Figure 6, where we cast $s$ vs. $k$, leaving $b$ momentarily  behind the scenes, crafting a parametric plot. Presenting the data as such, stripped of its hierarchical lattice origins, will permit us salient comparison to Euclidean DPRM results later. Even so,  within the figure inset, upper left, the explicit dependences, $s(b)$ and $k(b)$ are indicated, where we have chosen a logarithmic scale for the abscissa, in accordance with expectations regarding $d_{\rm eff}(b)$ for the $\diamondsuit$DPRM.  Hence, this inset reveals the behaviors of $s$ and $k$ as a function of the branching parameter $b$; i.e., effectively, dimensionality.  In particular, while $s$ and $k$ both vanish in the Gaussian limit ($b=1$), the skewness rises more rapidly than the kurtosis initially, scaling roughly as the square root.  By contrast, the kurtosis appears to have positive curvature at the inset origin, but
quickly straightens out to something essentially linear, though with slightly negative curvature beyond the curious (inflection?) point where the kurtosis and
skewness intersect, sharing the common value $s^\star$=$k^\star\hspace{-0.8mm}\approx$0.49375 at $b\hspace{-0.8mm}\approx$2.82826; recall, again, Fig 5. 
Of course, negative curvature in $k$ is expected, given the anticipated infinite $b$ limit; i.e., standard Gumbel, with $s_G$=1.14 and $k_G$=$2.4$. We emphasize that our measured $s$ and $k$ values are asymptotic, invoking the characteristic KPZ finite-time correction, which vanishes as $t^{-2\omega}$.  Returning to Figure 6 proper, one senses the possibility of a power-law relation,
$k\sim s^\rho$, between the $\diamondsuit$DPRM skewness and kurtosis; perhaps, {\it parabolic.}  This is investigated in a second inset, placed lower right, where we plot the running exponent $\rho_{\rm eff}$=$\partial{\rm ln}k/\partial{\rm ln}s$.  Indeed, for increasing ``dimensionality" (i.e., $b\uparrow$), as both $s$ and $k$ grow beyond the common intersection $s^*$=$k^*\approx\frac{1}{2}^-$, heading towards their respective standard Gumbel values, the index $\rho\approx 2^-$. By contrast, in the opposite limit, as $b$ heads towards unity, where the limit distribution becomes Gaussian with vanishing skewness and kurtosis, we see a different, quite distinct exponent $\rho\approx 1^+$. That said, it is manifest already that at $b$=2, where $s$=0.368, the index $\rho\approx1.79$, well on its way towards quadratic behavior. 

Finally, Figure 6 proper also includes individual data points associated with  our measured $(s,k)$ values  for the $d$+1 Euclidean KPZ/DPRM problem, in transverse dimensionalities $d$=0-6.  We note, in particular, that for unconstrained 3+1 DPRM extremal trajectories, with Gaussian distributed random energies on the sites \cite{HH13}, we find $s$=$k$=0.51$^-$; see discussion below.
Thus, somewhat serendipitously, it appears that 3+1 KPZ provides that instance in which the skewness and kurtosis slide past each other numerically; in fact, within the error bars, it is the same such value exhibited by the $b$=2.9 $\diamondsuit$DPRM. Of course, the coincidence of $s$ and $k$ hasn't the slightest mathematical significance. It is, rather, just an idiosyncratic feature of the 3+1 {\it flat} KPZ Class; another being the existence of a finite-temperature phase transition separating
entropic $(T>T_c)$ and extremal $ (T<T_c)$ DPRM wandering \cite{Golinelli}.  Nevertheless, that the coincident value is nearly the same, $s^*$=$k^*\approx\frac{1}{2}$, for {\it both} the Euclidean and hierarchical DPRM is, quite frankly, a little surprising.  Additionally, it raises many natural questions regarding the fundamental characteristics shared (and not...) of these similar, yet topologically distinct, extremal path lattice models.

\subsection{{\bf The Many-Dimensional DPRM \& Fate of $d$=$\infty$ KPZ}}

We comment further here regarding Figure 6 and associated results, as well as the higher-dimensional KPZ context within which they sit-

\noindent $\bullet$ Prior studies \cite{DG89,HH89,BD90,Roux91,MG08} of the $\diamondsuit$DPRM focused, initially at least, upon the exponent $\omega$ and had confirmed (one might say, in part, lamented...)
the slightly ``low" value $\omega$=0.299 found for $b$=$d_{\rm eff}^\diamondsuit$=2, given the exact result $\omega$=$\frac{1}{3},$ well-known for the 1+1 KPZ/DPRM problem.  A glance at Figure 5 insert, however, reveals that there is, of course, a value of the branching parameter for which this famous exponent is actually retrieved. Indeed, for $b$=1.69, we find $\omega$=0.3338$^-$ AND, quite surprisingly (this is the main point....), $s$=0.2947 and $k$=0.1956; i.e., values nearly coincident with TW-GOE results, $s_1$=0.2935 and $k_1$=0.1652, respectively. An additional pay-off comes when we examine the asymptotic scaling behavior of the first and second moments, extracting the $\diamondsuit$DPRM analog of the TW-GOE universal ratio $\langle\chi_1\rangle/\langle\chi_1^2\rangle^{1/2}=0.9515$,
which marks the mean-to-width characteristic of the underlying limit distribution. Here, we discover 0.86. In other words, the $b$=1.69 $\diamondsuit$DPRM PDF provides a reasonably good match to its 1+1 KPZ/TW-GOE cousin. Since $\omega$ is dead-on, the long TW Airy tail ($\eta=\frac{3}{2}$) is captured, along with the distribution's skewness and nearly, but not quite, its kurtosis. Agreed, the steep tail exponent, $\eta'=d_{\rm eff}\eta\approx2.64$, is a bit shy of the Tracy-Widom value 3 but, we note, with some irony, very much in line with existing heroic \cite{KBM,MG06}, though admittedly insuffucient, numerical estimates of this stubbornly difficult, intractable quantity. Hence, in a {\it zero-mean, unit-variance} rendering, the $b$=1.69 $\diamondsuit$DPRM PDF would sit comfortably amidst standard  1+1 KPZ class model simulation data (e.g., RSOS, PNG, etc.)  unless one plunged, at substantial numerical cost, deep into the tails.  Nevertheless, as we will discuss in detail momentarily, the clear numerical distinction 
between these $k$ values (esp., the fact that $k_{\rm b=1.69}>k_{\rm TW-GOE}$, despite the coincidence of the skewness $s$...) will be of great consequence.

\noindent $\bullet$ Recalling recent 2+1 DPRM results \cite{HH12}, which have isolated the higher-dimensional analog of KPZ/TW-GOE, measuring $s$=0.424 and $k$=0.346 for the generic {\it pt-plane} case, and pinned down the key index $\omega_{2+1}$=0.241 \cite{FT,KO} via a multi-model study \cite{HH13} of 2+1 KPZ {\it stationary-state} statistics, we notice quite similar values for the $b$=2.3 $\diamondsuit$DPRM, where the exponent, 0.274, might be a little high, but the skewness, 0.423, and kurtosis, 0.372, certainly close to the mark. Given our findings above for the $\diamondsuit$DPRM $sk$ relation, one cannot resist looking at Euclidean KPZ from this same vantage point. Indeed, naively fitting known 1+1 TW-GOE and 2+1 KPZ/DPRM \cite{HH12} values for the skewness and kurtosis to a simple power-law yields $k$=1.94$s^{2.01}$; i.e., a {\it quadratic} dependence. We encourage the engaged reader to dust off their calculator, check these numbers, and then ``compute" $k$ for TW-GUE $s$=0.2241. This is at least a little surprising, no?
Next, consider the 3d {\it pt-line} DPRM \cite{HH13}-Appendix, for which $(s,k)$=(0.408,0.322), as well as the earlier discussed 3d {\it pt-pt} DPRM/KPZ problem,
where $(s,k)$=(0.328,0.214).  In these instances, as well, the ratio $k/s^2$=2$^-$; in fact, eerily so.   Data points associated with these additional pt-pt and pt-line KPZ subclasses have thus been included in Figure 6, filling out the curve a bit.   

\noindent $\bullet$Curious about these matters, we have gone on to higher dimensional {\it hypercubic} DPRM, measuring for the 3+1 {\it flat} case: $\omega$=0.1868, $s$=0.508 and $k$=0.509, with system sizes $L$=500 and $\approx$10$^{11}$ data.  
For a distinct, but related polymer with random site energies drawn uniformly from the unit interval, so-called u5 DPRM \cite{HH12}, we find quite similar values $s$=$k$=0.515. 
As disclosed earlier, our main numerical observation regarding the 3+1 DPRM/KPZ problem is that already in {\it this} dimensionality, the limit distribution kurtosis has risen, numerically, to the level of the skewness. More to the point, however, we now supplement that arithmetic fact with our $sk$-relation ``prediction" $s^*$=$k^*$=1.94$^{-1}\approx 0.51$.  
For the 4+1 Gaussian DPRM, we find $\omega$=0.152$\pm$0.004, $s$=0.577 and $k$=0.688, respectively, for $L$=125. Things become progressively more challenging in higher dimensions, but for the 5+1 DPRM: $(s,k)$=(0.596,0.73), and 6+1: $(s,k)$=(0.623,0.805). {\it These findings, when cast in terms of an $sk$-plot, suggest that the $\diamondsuit$DPRM is quite distinct from its Euclidean cousin.}  We have illustrated this explicitly for the $b$=1.69 hierarchical lattice, whose limit distribution possesses $\omega$=$\frac{1}{3}$, shares the TW-GOE skewness, but has a kurtosis, $k$=0.1956(1), definitively {\it larger} than the Tracy-Widom value $k_1$=0.1652. In other words, for small skewness, the $sk$-curve of the $\diamondsuit$DPRM rides {\it well above} the Euclidean DPRM data; however, the respective $sk$-traces eventually cross     
near the transcendent value $s^*$=$k^*\approx\frac{1}{2}$. Since it is known that asymptotically, as $b\rightarrow\infty$, the $\diamondsuit$DPRM limit distribution 
converges to standard Gumbel- i.e., the point $(s_G,k_G)$, upper right corner Figure 6, proper- this suggests that $d$=$\infty$ KPZ, in its DPRM manifestation,
{\it is NOT simply Gumbel.} The $sk$-trace of the many-dimensional DPRM is, at the start, distinct, but then rises precipitously  at a different rate. Indeed, this suggestion is consistent with independent work \cite{Dean} investigating directed polymers on Cayley trees \cite{DS88}, effectively an infinite dimensional implementation of the KPZ problem.

\noindent $\bullet$ {\it Parting gesture:}  On a rather different note- Folding together in one hand the KK conjecture \cite{KK89}, $\omega=1/(d+2)$, for the $d$+1 DPRM/KPZ problem, Zhang's formula for the  TW Airy tail,  
$\eta=1/(1-\omega),$ along with the MG tail-exponent relation leads to the hopelessly seductive, simple result  that $\eta=(d+2)/(d+1)$ and $\eta\prime=d+2$ for {\it Euclidean} lattices; in other words, $(\eta,\eta\prime)=(\frac{3}{2},3)$ known  rigorously for the  1+1 KPZ/DPRM TW limit distribution, where $d$=1, while for the 2+1 DPRM:
$(\eta,\eta\prime)=(\frac{4}{3},4),$ and in 3+1: $(\eta,\eta\prime)_{3+1}=(\frac{5}{4},5)$, etc. This is almost certainly not true, but frankly, nearly irrefutable numerically \cite{MG06}  (even admitting the loss of KK....) and,
we must confess, an irresistible, if only convenient, rule-of-thumb.  


\section{The Future}
We close with a passing observation, then a rallying call.  Firstly, for a strict {\it quadratic} dependence, whereby $\rho\equiv2,$ the quantity $k/s^\rho\rightarrow\langle\delta h^4\rangle_c\langle\delta h^2\rangle_c/\langle\delta h^3\rangle_c^2$, essentially the Derrida-Appert ratio \cite{DA99}.   From this perspective, our work here reveals that the Derrida-Appert ratio is, despite quite a broad range of underlying $s$ and $k$ values,  {\it nearly constant for the entire Euclidean KPZ problem,} across many dimensionalities, as well as geometric subclasses: pt-pt, pt-line, pt-plane, etc. Indeed, the variation of this quantity is quite modest,
running from TW-GUE: 1.86 to 6+1 DPRM: 2.07.  Clearly, $\rho\ne2$ exactly.  Nevertheless, we wonder whether a  tightly-knit family of limit distributions may someday be implicated in these numerical conspiracies, or otherwise explain the findings  of Figure 6. 
Until then, we encourage mathematically-inclined colleagues to revisit the $\diamondsuit$DPRM and related matters.  Any rigorous statement (e.g., a bound) regarding the tail exponents of its  $b-$parametrized limit distributions would be a great boon, certainly. Of course, an exact, full solution for a single $b$-value wouldn't be turned away either; in sum, any additional wisdom regarding the $\diamondsuit$DPRM would be much appreciated.

\begin{acknowledgements}
The authors would like to express their gratitude to Herbert Spohn for his many years of inspired work, wisdom, and stamina on behalf of the KPZ cause. 
Thanks, too, to Joel Lebowitz for keeping the statistical mechanical fire well-lit through the generations.
This work is supported in part by KAKENHI (No. 25707033 from JSPS and No. 25103004 ``Fluctuation \& Structure'' from MEXT in Japan).
\end{acknowledgements}


\begin{thebibliography}{}

\bibitem{KPZ} Kardar, M., Parisi, G., Zhang, Y.-C.: Dynamic scaling of growing interfaces. Phys. Rev. Lett. {\bf 56}, 889 (1986)
\bibitem{KMHH} Krug, J., Meakin, P., Halpin-Healy, T.:  Amplitude universality for driven interfaces and directed polymers in random media. Phys. Rev. A{\bf 45}, 638 (1992)
\bibitem{S12} Spohn, H.:  KPZ scaling theory and the semi-discrete directed polymer. arXiv: 1201.0645
\bibitem{KM} Krug, J., Meakin, P.: Universal finite-size effects in the rate of growth processes. J. Phys. A{\bf 23}, L987 (1990)
\bibitem {HHF} Huse, D. A., Henley, C. L., Fisher, D. S.: Forced Burgers equation, exact exponent, fluctuation-dissipation theorem, Phys. Rev. Lett. {\bf 55}, 2924 (1985)
\bibitem{Dhar} Dhar, D.: An exactly solved model for interface growth. Phase Transitions {\bf 9}, 51 (1987)
\bibitem{Gwa1} Gwa, L.-H., Spohn, H.: Six-vertex model, roughened surfaces, and an asymmetric spin hamiltonian. Phys. Rev. Lett. {\bf 68}, 725 (1992)
\bibitem{Gwa2} Gwa, L.-H., Spohn, H.: Bethe solution for the dynamical-scaling exponent of the noisy Burgers equation. Phys. Rev. A{\bf 46}, 844 (1992)
\bibitem{KB} Kardar, M.: Replica Bethe ansatz studies of two-dimensional interfaces with quenched random impurities.  Nucl. Phys. B{\bf 290}, 582 (1987)
\bibitem{DNR} Kardar, M., Nelson, D. R.: Commensurate-incommensurate transitions with quenched disorder. Phys. Rev. Lett. {\bf 55}, 1157 (1985)
\bibitem{IS88} Imbrie, J. Z., Spencer, T.: Diffusion of directed polymers in a random environment. J. Stat. Phys. {\bf 52}, 609 (1988)
\bibitem{Dot} Doty, C. A., Kosterlitz, J. M.: Exact dynamical exponent at the Kardar-Parisi-Zhang roughening transition.
Phys. Rev. Lett. {\bf 69}, 1979  (1992)
\bibitem{Medina} Medina, E., Hwa, T., Kardar, M., Zhang, Y.-C.: Burgers equation with correlated noise: Renormalization-group analysis and applications to directed polymers and interface growth. Phys. Rev. A {\bf 39}, 3053 (1989)
\bibitem{2loop} Frey, E., T\"auber: Two-loop renormalization-group analysis of the Burgers-"Kardar-Parisi-Zhang equation. Phys. Rev. E {\bf 50}, 1024 (1994) 
\bibitem{BKS} van Beijeren, H., Kutner, R., Spohn, H.:  Excess noise for driven diffusive systems. Phys. Rev. Lett. {\bf 54}, 2026 (1985)
\bibitem{MCRG} Frey, E., T\"auber, U., Hwa, T.: Mode-coupling and renormalization group results for the noisy Burgers equation. Phys. Rev. E {\bf 53}, 4424 (1996)
\bibitem {HH85} Huse, D. A., Henley, C. L.: Pinning and roughening of domain walls in Ising systems due to random impurities, Phys. Rev. Lett. {\bf 54}, 2708 (1985)
\bibitem {MK85} Kardar, M.: Roughening by impurities at finite temperatures, Phys. Rev. Lett. {\bf 55}, 2923 (1985)
\bibitem{KK89} Kim, J.-M., Kosterlitz, M.: Growth in a restricted solid-on-solid model. Phys. Rev. Lett. {\bf 62}, 2289 (1989)
\bibitem{FT} Forrest, B., Tang, L.-H.:  Surface roughening in a hypercube-stacking model. Phys. Rev. Lett. {\bf 64}, 1405 (1990)
\bibitem{AF90} Amar, J. A., Family, F.: Numerical solution of a continuum equation for interface growth in 2+1 dimensions. Phys. Rev. A {\bf 41}, 3399 (1990)
\bibitem{FT2} Tang, L.-H., Forrest, B., Wolf, D. E.: Kinetic surface roughening. II. Hypercube-stacking models. Phys. Rev. A {\bf 45}, 7162 (1992)
\bibitem{MW94} Moser, K., Wolf, D.: Vectorized and parallel simulations of the KPZ equation in 3+1 dimensions. J. Phys. A {\bf 27}, 4049 (1994).
\bibitem{HF} Hwa, T., Frey, E.: Exact scaling function of interface growth dynamics. Phys. Rev. E {\bf 44}, R7873 (1991)
\bibitem{LeiHan} Tang, L.-H.: Steady-state scaling function of the (1 + 1)-dimensional single-step model. J. Stat. Phys. {\bf 67}, 819 (1992)
\bibitem{KBM} Kim, J. M., Moore, M. A., Bray, A.-J.: Zero-temperature directed polymers in a random potential.  Phys. Rev. A {\bf 44}, 2345 (1991) 
\bibitem{HH91} Halpin-Healy, T.:  Directed polymers in random media: Probability distributions. Phys. Rev. A {\bf 44}, R3415 (1991)
\bibitem{PM93} Meakin, P.: The growth of rough surfaces and interfaces. Phys. Rep. {\bf 235}, 189 (1993)
\bibitem{HHZ} Halpin-Healy, T., Zhang, Y.-C.:  Kinetic roughening phenomena, stochastic growth, directed polymers and all that. Aspects of multidisciplinary statistical mechanics. Phys. Rep. {\bf 254}, 215 (1995)
\bibitem{K97} Krug, J.: Origins of scale invariance in growth processes. Adv. Phys. {\bf 46}, 139 (1997)
\bibitem{FFF97} Maunuksela, J.: Kinetic roughening in the slow combustion of paper. Phys. Rev. Lett. {\bf 79}, 1515 (1997); earlier-
J. Zhang: Modeling forest fire by paper-burning experiment. Physica A {\bf 189}, 383 (1992)
\bibitem{BG} Bertini, L., Giacomin, G.: Stochastic Burgers and KPZ equations from particle systems. Comm. Math. Phys. {\bf 183}, 571-607 (1997)
\bibitem{Gunter} Sch\"utz, G. M.: Duality relations for asymmetric exclusion processes. J. Stat. Phys. {\bf 86} 1265 (1997)
\bibitem{LK} L\"assig, M., Kinzelbach, H.: Upper critical dimension of the Kardar-Parisi-Zhang equation. Phys. Rev. Lett. {\bf 78}, 903 (1997)
\bibitem{Lassig} L\"assig, M.: Quantized scaling of growing surfaces. Phys. Rev. Lett. {\bf 80}, 2366 (1998); for recent taste of conformal KPZ- Cao, X., Rosso, A., Santachiara, R.: Conformal invariance of loop ensembles under KPZ dynamics. arXiv: 1506.03291. Also, w/ SLE- Saberi, A. A. {\it et al.}: Classification of (2+1)-dimensional growing surfaces using Schramm-Loewner evolution. Phys. Rev. E {\bf 82}, 020101 (2010); as well- Phys. Rev. E {\bf 77}, 051607 (2008), Phys. Rev. Lett. {\bf 100}, 044504 (2008)
\bibitem{TAP98} Ala-Nissila, T.: Upper critical dimension of the Kardar-Parisi-Zhang equation. Phys. Rev. Lett. {\bf 80}, 887 (1998)
\bibitem{JMK98} Kim, J. M.: Phase transition of the KPZ equation in four substrate dimensions. Phys. Rev. Lett. {\bf 80}, 888 (1998)
\bibitem{Roma} Castellano, C., Marsili, M., Pietronero, L.: Nonperturbative renormalization of the Kardar-Parisi-Zhang growth dynamics. Phys. Rev. Lett. {\bf 80}, 3527 (1998)
\bibitem{Parisi} Marinari, E., Pagnani, A., Parisi, G.:  Critical exponents of the KPZ equation via multi-surface coding numerical simulations. J. Phys. A {\bf 33}, 8181 (2000)
\bibitem{HH90} Halpin-Healy, T.: Disorder-induced roughening of diverse manifolds. Phys. Rev. A {\bf 42}, 711 (1990)
\bibitem{FTJ} Frey, E., T\"auber, U., Janssen, H. K.: Scaling regimes and critical dimensions in the Kardar-Parisi-Zhang problem. Europhys. Lett. {\bf 47}, 14 (1999)
\bibitem{PS12} Perlsman, E., Schwartz, M.: UCD of the KPZ equation. Phys. Rev. E {\bf 85}, 050103 (2012)
\bibitem{Parisi13} Pagnani, A., Parisi, G.: Multisurface coding simulations of the RSOS model in four dimensions. Phys. Rev. E {\bf 87}, 010102 (2013)
\bibitem{JMK13} Kim, J. M., Kim, S.-W.: RSOS model with a proper restriction parameter N in 4+1 dimensions. Phys. Rev. E {\bf 88}, 034102 (2013)
\bibitem{Alves14} Alves, S. G., Oliveira, T. J., Ferreira, S. C.: Universality of fluctuations in the Kardar-Parisi-Zhang class in high dimensions and its upper critical dimension. Phys. Rev. E {\bf 90}, 020103 (2014)
\bibitem{glassy} Moore, M. A., Blum, T., Doherty, J. P., Marsili, M., Bouchaud, J.-P., Claudin, P.:  Glassy solutions of the Kardar-Parisi-Zhang equation. Phys. Rev. Lett. {\bf 74,} 4257 (1995);
also- J. P. Doherty, {\it et al.}: O(N) Generalization of the KPZ equation. Phys. Rev. Lett. {\bf 72,} 2041 (1994).
Henkel, M. and Durang, X.: Spherical model of growing interfaces. JSTAT (2015) P05022
\bibitem{CCDW} Can\'et, L., Chat\'e, H., Delamotte, B., Wschebor, N.: Nonperturbative renormalization group for the KPZ Equation. Phys. Rev. Lett. {\bf 104}, 150601 (2010)

\bibitem{PS1} Pr\"{a}hofer, M., Spohn, H., Universal distributions for growth processes in 1+1 dimensions and random matrices. Phys. Rev. Lett. {\bf 84}, 4882 (2000)
\bibitem{KJ} Johansson, K.: Shape fluctuations and random matrices. Commun. Math. Phys. {\bf 209}, 437 (2000) 
\bibitem{TW} Tracy, C., Widom, H.: Level-spacing distributions and the Airy kernel.  Commun. Math. Phys. {\bf 159}, 151 (1994);
{\it ibid,} {\bf 177}, 727 (1996)
\bibitem{BDJ} Baik, J., Deift, P., Johansson, K.: On the distribution of the length of the longest increasing subsequence of random permutations. J. Am. Math. Soc. {\bf 12}, 1119 (1999)
\bibitem{AD99} Aldous, D., Diaconis, P.:  Longest increasing subsequences: from patience sorting to the Baik-Deift-Johansson theorem. Bull. Amer. Math. Soc. {\bf36}, 413 (1999)
\bibitem{Andrei00}  Okounkov, A.: Random matrices and random permutations. Int. Math. Res. Not. {\bf 2000}, 1043 (2000) 
\bibitem{OR99} Odlyzko, A. M., Rains, E. M., ATT Bell Labs Technical Report (1999)
\bibitem{BB68} Baer, R. M., Brock, P.: Natural sorting over permutation spaces. Math. Comput. {\bf 22}, 385 (1968)
\bibitem{BR} Baik, J., Rains, E. M.: Limiting distributions for a polynuclear growth model with external sources. J. Stat. Phys. {\bf 100}, 523 (2000)
\bibitem{PS2} Pr\"ahofer, M., Spohn, H.: Scale invariance of the PNG droplet and the Airy process. J. Stat. Phys. {\bf 108}, 1071  (2002)
\bibitem{FFF1} Myllys, M., Maunuksela, J., Alava, M., Ala-Nissila, T., Merikoski, J., Timonen, J.: Kinetic roughening
in slow combustion of paper. Phys. Rev. E {\bf 64}, 036101 (2001)
\bibitem{FFF3} Myllys, M.: Effect of a columnar defect on the shape of slow-combustion fronts. Phys. Rev. E {\bf 68}, 051103 (2003); also, J. Krug and L.-H. Tang: Disorder-induced unbinding in confined geometries.
Phys. Rev. E {\bf 50}, 104 (1994).
\bibitem{FFF5}  Miettinen, L., Myllys, M., Merikoski, J., Timonen, J.: Experimental determination of KPZ height-fluctuation distributions. Eur. Phys. J. B {\bf 46}, 55 (2005)
\bibitem{Colaiori} Colaiori, F., Moore, M. A.: Upper critical dimension, dynamic exponent, and scaling functions in the mode-coupling theory for the KPZ equation. Phys. Rev. Lett. {\bf 86}, 3946 (2001)
\bibitem{HF05} Fogedby, H.: Localized growth modes, dynamic textures, and UCD for the KPZ Equation in the weak-noise limit. Phys. Rev. Lett. {\bf 94}, 195702 (2005)
\bibitem{HF06} Fogedby, H.: Kardar-Parisi-Zhang equation in the weak noise limit: Pattern formation and upper critical dimension. Phys. Rev. E {\bf 73}, 031104 (2006) 
\bibitem{GP} Palasantzas, G.:  Roughening aspects of room temperature vapor deposited oligomer thin films onto Si substrates. Surf. Sci {\bf 507}, 357 (2002)
\bibitem{Luna} Halpin-Healy, T.,  Lin, Y.: Universal aspects of curved, flat and stationary-state KPZ statistics. Phys. Rev. E {\bf 89}, 010103 (2014)
\bibitem{MN} Majumdar, S. N., Nechaev, S.: Anisotropic ballistic deposition model with links to the Ulam problem and the Tracy-Widom distribution. Phys. Rev. E {\bf 69}, 011103 (2004)
\bibitem{Toom} Barkema, G. T., Ferrari, P. L., Lebowitz, J. L., Spohn., H.: Kardar-Parisi-Zhang universality class and the anchored Toom interface. Phys. Rev. E {\bf 90}, 042116 (2014)

\bibitem{KK} Kriecherbauer, T., Krug, J.: A pedestrian's view on interacting particle systems, KPZ universality and random matrices.  J. Phys. A. {\bf 43}, 403001 (2010)
\bibitem{IC} Corwin, I.:  The KPZ equation and universality class. Random Matrices: Theory Appl. {\bf 1}, 1130001 (2012) 
\bibitem{SS10} Sasamoto, T., Spohn, H.: One-dimensional Kardar-Parisi-Zhang equation: An exact solution and its universality,  Phys. Rev. Lett. {\bf 104}, 230602 (2010)
\bibitem{Amir} Amir, G., Corwin, I., Quastel, J.: Probability distribution of the free energy of the continuum directed random polymer in 1 + 1 dimensions.  Commun. Pure Appl. Math {\bf 64}, 466 (2011)
\bibitem{CDR} Calabrese, P., Le Doussal, P., Rosso, A.: Free-energy distribution of the directed polymer at high temperature. Europhys. Lett. {\bf 90}, 20002 (2010)
\bibitem{Dotsenko} Dotsenko, V.: Bethe ansatz derivation of the Tracy-Widom distribution for one-dimensional directed polymers. Europhys. Lett. {\bf 90}, 20003 (2010)
\bibitem{TW09} Tracy, C. A., Widom, H.: Asymptotics in ASEP with step initial condition. Commun. Math. Phys. {\bf 290,} 129 (2009).
\bibitem{Calabrese} Calabrese, P., Le Doussal, P.: Exact solution for the Kardar-Parisi-Zhang equation with flat initial conditions.
Phys. Rev. Lett. {\bf 106}, 250603 (2011); JSTAT (2012) P06001
\bibitem{TG12}  Gueudr\'e, T., Le Doussal, P.: Directed polymer near a hard wall and KPZ equation in the half-space. EPL {\bf 100}, 26006 (2012)
\bibitem{IS} Imamura, T., Sasamoto, T.:  Exact solution for the stationary Kardar-Parisi-Zhang equation. Phys. Rev. Lett. {\bf 108}, 190603 (2012)
\bibitem{BC} Borodin, A., Corwin, I., Ferrari, P. L., Vet\H{o}, B.: Height fluctuations for the stationary KPZ equation. arXiv:1407.6977. 
\bibitem{TS10} Takeuchi, K. A., Sano, M.:  Universal fluctuations of growing interfaces: Evidence in turbulent liquid crystals. Phys. Rev. Lett. {\bf 104}, 230601 (2010)
\bibitem{TS11} Takeuchi, K. A., Sano, M., Sasamoto, T., Spohn, H.: Growing interfaces uncover universal fluctuations behind scale invariance.  Sci. Rep. (Nature) {\bf 1}, 34 (2011)
\bibitem{TS12} Takeuchi, K. A., Sano, M.: Evidence for geometry-dependent universal fluctuations of KPZ interfaces in liquid-crystal turbulence. J. Stat. Phys. {\bf 147}, 853-890 (2012)
\bibitem{KT13} Takeuchi, K. A.: Crossover from growing to stationary interfaces in the Kardar-Parisi-Zhang class. Phys. Rev. Lett. {\bf 110}, 210604 (2013)
\bibitem{KCW12} Kloss, T., Canet, L., Wschebor, N.: Nonperturbative renormalization group for the stationary Kardar-Parisi-Zhang equation: Scaling functions and amplitude ratios in 1+1, 2+1, and 3+1 dimensions. Phys. Rev. E{\bf 86}, 051124 (2012); see, esp., section IV-E.
\bibitem{HH12} Halpin-Healy, T.:   (2+1)-Dimensional directed polymer in a random medium: Scaling phenomena and universal distributions. Phys. Rev. Lett. {\bf 109}, 170602 (2012)
\bibitem{HH13} Halpin-Healy, T.: Extremal paths, the stochastic heat equation, and the 3d KPZ universality class. Phys. Rev. E {\bf 88}, 042118 (2013); {\it ibid,} {\bf 88}, 069903 (2013)
\bibitem{SF13} Oliveira, T. J., Alves, S. G., Ferreira, S. C.:  Kardar-Parisi-Zhang universality class in (2+1) dimensions: Universal geometry-dependent distributions and finite-time corrections. Phys. Rev. E {\bf 87}, 040102 (2013)
\bibitem{MP03} Pr\"ahofer, M.: Stochastic Surface Growth. Ludwig-Maximilians-Universitait, M\"unchen (2003)
\bibitem{Dutch} Halpin-Healy, T., Palasantzas, G.: Universal correlators \& distributions as experimental signatures of (2 + 1)-dimensional Kardar-Parisi-Zhang growth. EPL {\bf 105}, 50001 (2014)
\bibitem{Carrasco} Carrasco, I. S. S., Takeuchi, K. A., Ferreira, S. C., Oliveira, T. J.: Interface fluctuations for deposition on enlarging flat substrates, New J. Phys. {\bf 16}, 123057 (2014)
\bibitem{Almeida} Almeida, R. A. L., Ferreira, S. O., Oliveira, T. J., Aarao Reis, F. D. A.:  Universal fluctuations in the growth of semiconductor thin films. Phys. Rev. B {\bf 89}, 045309 (2014)
\bibitem{MH} Hairer, M.: Solving the KPZ equation. Ann. Math. {\bf 178}, 559 (2013)
\bibitem{TS} Sepp\"al\"ainen, T.: Scaling for a one-dimensional directed polymer with boundary conditions. Ann. Probab. {\bf 40}, 19 (2012)
\bibitem{OC} O`Connell, N.: Directed polymers and the quantum Toda lattice. Ann. Probab. {\bf 40} 437 (2012)
\bibitem{B1} Borodin, A., Corwin, I.: Macdonald processes. Prob. Th. Rel. Fields {\bf 158}, 225 (2014)
\bibitem{B2} Borodin, A., Corwin, I., Sasamoto, T.: From duality to determinants for q-TASEP and ASEP. Ann. Probab. {\bf 42}, 2314 (2014)
\bibitem{PC14} Calabrese, P., Kormos, M., Le Doussal, P.: From the sine-Gordon field theory to the Kardar-Parisi-Zhang growth equation. EPL {\bf 107}, 10011 (2014)
\bibitem{Hahn} Barraquand, G., Corwin, I.: The q-Hahn asymmetric exclusion process. arXiv:1501.03445
\bibitem{Petrov} Corwin, I., Petrov, L.: Stochastic higher spin vertex models on the line. arXiv:1502.07374
\bibitem{DSD15} Dean, D. S., Le Doussal, P., Majumdar, S. N., Schehr, G. Finite-temperature free fermions and the KPZ equation at finite time. Phys. Rev. Lett. {\bf 114}, 110402 (2015)
\bibitem{KJ15} Johansson, K.: Two-time distribution in Brownian directed percolation. arXiv:1502.00941

\bibitem{Maritan} Maritan, A., Toigo, F., Koplik, J., Banavar, J. R., Dynamics of growing interfaces.  Phys. Rev. Lett. {\bf 69}, 3193 (1992)  
\bibitem{trees} Batchelor, M. T., Henry, B. I., Watt, S. D.: Continuum model for radial interface growth. Physica A {\bf 260}, 11 (1998)
\bibitem{Singha} Singha, S. B.: Persistence of surface fluctuations in radially growing surfaces. J. Stat. Mech. (2005) P08006
\bibitem{Mas} Masoudi, A. A.: Statistical analysis of radial interface growth. JSTAT {\bf 2012}, L02001 (2012)
\bibitem{SS11} Rodriguez-Laguna, J., Santalla, S. N.,  Cuerno, R.:  Intrinsic geometry approach to surface kinetic roughening. J. Stat. Mech. (2011) P05032
\bibitem{SS14} Santalla, S. N., Rodriguez-Laguna, J., Cuerno, R.: The circular Kardar-Parisi-Zhang equation as an inflating, self-avoiding ring polymer. 
Phys. Rev. E {\bf 89}, 010401 (2014)
\bibitem{SS15} Santalla, S. N., Rodriguez-Laguna, J., LaGatta, T., Cuerno, R.:  Random geometry and the Kardar-Parisi-Zhang universality class. New J. Phys. {\bf 17}, 033018 (2015)
\bibitem{Alves11} Alves, S. G., Oliveira, T. J., Ferreira, S. C.: Universal fluctuations in radial growth models belonging to the KPZ universality class. Europhys. Lett. {\bf 96}, 48003 (2011)
\bibitem{Kaz12} Takeuchi, K. A.: Statistics of circular interface fluctuations in an off-lattice Eden model.  J. Stat. Mech. {\bf 2012}, P05007 (2012)
\bibitem{SA13} Alves, S. G., Oliveira, T. J., Ferreira, S. C.: Non-universal parameters, corrections and universality in Kardar-Parisi-Zhang growth. JSTAT {\bf 2013}, P05007 (2013)
\bibitem{Bourne} Bornemann, F.: On the numerical evaluation of Fredholm determinants. Math. Comput. {\bf 79}, 871–915 (2010)
\bibitem{Racz} R\'acz, Z., Plischke, M.: Width distribution for (2+1)-dimensional growth and deposition processes. Phys. Rev. E {\bf 50}, 3530 (1994)
\bibitem{Foltin} Foltin, G., Oerding, K., R\'acz, Z., Workman, R. L., Zia, R. K. P.: Width-distribution for random-walk interfaces.  Phys. Rev. E {\bf 50}, R639 (1994)
\bibitem{EW} Edwards, S. F., Wilkinson, D. R.:  The surface statistics of a granular aggregate. Proc. R. Soc. London Ser. A {\bf 381}, 17 (1982).
\bibitem{Antal} Antal, T., Droz, M., Gy\H{o}rgyi, G., R\'acz, Z.: Roughness distributions for 1/f$^\alpha$ signals. Phys. Rev. E {\bf 65}, 046140 (2002) 
\bibitem{Santa} Santachiara, R., Rosso, A., Krauth, W.: Universal width distribution in non-Markovian gaussian processes, JSTAT 2007, P02009 (2007)
\bibitem{Shapir} Raychaudhauri, S., Cranston, M., Przybyla, C., Shapir, Y.:  Maximal height scaling of kinetically growing surfaces. Phys. Rev. Lett., {\bf 87}, 136101 (2001)
\bibitem{Comtet} Majumdar, S. N., Comtet, A.: Exact maximal height distribution of fluctuating interfaces. Phys. Rev. Lett. {\bf 92}, 225501 (2004)
\bibitem{Comtet2} Majumdar, S. N., Comtet, A.: Airy distribution function: From the area under a Brownian excursion to the maximal height of fluctuating interfaces. J. Stat. Phys., {\bf 119}, 777 (2005)
\bibitem{GS} Schehr, G., Majumdar, S.: Universal asymptotic statistics of maximal relative height in one-dimensional solid-on-solid models. 
Phys. Rev. E{\bf 73}, 056103 (2006)
\bibitem{Gyorgyi2} Gy\H{o}rgyi, G., Moloney, N. R., Ozog\'any, K., R\'acz, Z.: Maximal height statistics for 1/$f^a$ signals.  Phys. Rev. E {\bf 75}, 021123 (2007) 
\bibitem{Rambeau2} Rambeau, J., Bustingorry, S., Kolton, A. B., Schehr, G.: MRH of elastic interfaces in random media,  Phys. Rev. E {\bf 84}, 041131 (2011)
\bibitem{DSL} Lee, D.-S.: Distribution of extremes in the fluctuations of two-dimensional equilibrium interfaces.  Phys. Rev. Lett. {\bf 95}, 150601 (2005)
\bibitem{KO} Kelling, J., \'Odor, G.:  Extremely large-scale simulation of a Kardar-Parisi-Zhang model using graphics cards. Phys. Rev. E {\bf 84}, 061150 (2011)
\bibitem{DG89} Derrida, B., Griffiths, R.: Directed polymers on disordered hierarchical lattices.  Europhys. Lett. {\bf 8}, 111 (1989); also-
Cook, J., Derrida, B.: Polymers on disordered hierarchical lattices: A nonlinear combination of random variables. J. Stat. Phys. {\bf 57}, 89 (1989)
\bibitem{HH89} Halpin-Healy, T.: Comment- Growth in a restricted solid-on-solid model. Phys. Rev. Lett. {\bf 63}, 917 (1989)
\bibitem{BD90} Derrida, B.: Directed polymers in a random medium: Physica A {\bf 163}, 71 (1990)
\bibitem{Roux91} Roux, S., Hansen, A., da Silva, L., Lucena, L., Pandey, R.: Minimal path on the hierarchical diamond lattice.
J. Stat. Phys. {\bf 65}, 183 (1991)
\bibitem{MG08} Monthus, C., Garel, T.: Disorder-dominated phases of random systems: relations between the tail exponents and scaling exponents.
J. Stat. Mech. {\bf 2008}, P01008 (2008)
\bibitem{Gumbel} Gumbel, E.J.: Statistics of Extremes. Columbia Univ. Press. New York (1958). Republished by Dover, New York (2004)
\bibitem{FF} Ferrari, P.L., Frings, R.: Finite-time corrections in KPZ growth models. J. Stat. Phys. {\bf 144}, 1123 (2011);
Oliveira, T. J., Ferreira, S. C., Alves, S. G.:  Universal fluctuations in KPZ growth on one-dimensional flat substrates. Phys. Rev. E {\bf 85}, 010601 (2012)
\bibitem{Golinelli} Derrida, B., Golinelli, O.: Thermal properties of directed polymers in random media. Phys. Rev. A {\bf 41}, 4160 (1990); too-
Monthus, C., Garel, T.: Numerical study of the directed polymer in a 3+1 dimensional random medium. Eur. Phys. J. B {\bf 53}, 39 (2006)
\bibitem{MG06} Monthus, C., Garel, T.: Probing the tails of the ground-state energy distribution for the directed polymer in a random medium of dimension d=1,2,3 via a Monte Carlo procedure in the disorder. Phys. Rev. E {\bf 74}, 051109 (2006)
\bibitem{Dean} Dean, D. S., Majumdar, S. N.: Extreme-value statistics of hierarchically correlated variables deviation from Gumbel statistics and anomalous persistence. Phys. Rev. E {\bf 64}, 046121 (2001)
\bibitem{DS88} Derrida, B., Spohn, H.: Polymers on disordered trees, spin glasses, and traveling waves. J. Stat. Phys. {\bf 51}, 817 (1988)
\bibitem{DA99} Derrida, B., Appert, C.: Universal large deviation function of the KPZ equation in one dimension. J. Stat. Phys. {\bf 94}, 1 (1999)


\end{thebibliography}


\end{document}